\documentclass[twocolumn,secnumarabic,amssymb, nobibnotes, aps, prl,superscriptaddress,notitlepage]{revtex4}
\usepackage{amssymb,amsfonts,amstext,mathrsfs}
\usepackage{amsthm}
\usepackage{dcolumn}
\usepackage{bm}
\usepackage{graphicx,breakurl}
\usepackage{amsmath}
\usepackage{color,setspace}
\usepackage{lipsum}
\usepackage[T1]{fontenc}
\usepackage[normalem]{ulem}

\usepackage{algorithm}  
\usepackage{algorithmicx}  
\usepackage{algpseudocode}  
\usepackage{booktabs}

\usepackage[colorlinks,
           linkcolor=blue,
            anchorcolor=blue,
             citecolor=blue
            ]{hyperref}
            
\begin{document}

\title{Experimental demonstration of adversarial examples in learning topological phases}
\date{\today}
\author{Huili Zhang}\thanks{These authors contributed equally to this work.}
\author{Si Jiang }\thanks{These authors contributed equally to this work.}
\author{Xin Wang}\thanks{These authors contributed equally to this work.}
\author{Wengang Zhang}\author{Xianzhi Huang}\author{Xiaolong Ouyang}\author{Yefei Yu}\author{Yanqing Liu}
\affiliation{Center for Quantum Information, IIIS, Tsinghua University, Beijing 100084, P. R. China}

\author{Dong-Ling Deng}\email{dldeng@tsinghua.edu.cn}
\affiliation{Center for Quantum Information, IIIS, Tsinghua University, Beijing 100084, P. R. China}
\affiliation{Shanghai Qi Zhi Institute, 41th Floor, AI Tower, No. 701 Yunjin Road, Xuhui District, Shanghai 200232, China}

\author{L.-M. Duan}\email{lmduan@tsinghua.edu.cn}
\affiliation{Center for Quantum Information, IIIS, Tsinghua University, Beijing 100084, P. R. China}

\begin{abstract}
\textbf{Classification and identification of different phases and the transitions between them is a central task in condensed matter physics. Machine learning, which has achieved dramatic success in a wide range of applications, holds the promise to bring unprecedented perspectives for this challenging task. However, despite the exciting progress made along this direction, the reliability of  machine-learning approaches likewise demands further investigation. Here, with the nitrogen-vacancy center platform, we report the first proof-of-principle experimental demonstration of adversarial examples in learning topological phases. We show that, after adding a tiny amount of carefully-designed perturbations, the experimentally observed adversarial examples can successfully deceive a splendid phase classifier, whose prediction accuracy is larger than $99.2\%$ on legitimate samples, with a notably high confidence. Our results explicitly showcase the crucial vulnerability aspect of applying machine learning techniques in classifying phases of matter, which provides an indispensable guide for future studies in this interdisciplinary field.}
\end{abstract}
\maketitle
Machine learning, or more generally speaking artificial intelligence, is currently taking a technological revolution to modern society and becoming a powerful tool for fundamental research in multiple disciplines \cite{LeCun2015Deep, Jordan2015Machine}. Recently, machine learning has been adopted to solve challenges in condensed matter physics \cite{carleo2019machine, Sarma2019Machine, Dunjko2018Machine}, and in particular, to classify phases of matter and identify phase transitions \cite{li2018machine,Wang2016Discovering,Carrasquilla2017Machine,vanNieuwenburg2017Learning,
Chng2017Machine,Zhang2017Quantum}. Within this vein, both supervised \cite{Zhang2017Quantum,Zhang2017Machine,Zhang2018Machine} and unsupervised learning  \cite{Wang2016Discovering,rodriguez2019identifying, Yu2021Unsupervised, Scheurer2020Unsupervised, Long2020Unsupervised, Lidiak2020Unsupervised} 
methods have been applied, enabling identifying different phases directly from raw data of local observables, such as spin textures and local correlations \cite{Zhang2017Quantum,Zhang2018Machine,Zhang2017Machine}. In addition, pioneering experiments have also been carried out with different platforms \cite{lian2019machine, Rem2019Identifying, Bohrdt2019Classifying, Zhang2019Machine}, including electron spins in nitrogen-vacancy (NV) centers in diamond \cite{lian2019machine}, cold atoms in optical lattices \cite{Rem2019Identifying, Bohrdt2019Classifying}, and doped $\rm{CuO_2}$ \cite{Zhang2019Machine}, showing  unparalleled potentials of machine learning approaches compared to traditional means.

\begin{figure*}[htbp]
  \centering
  \includegraphics[width=18cm]{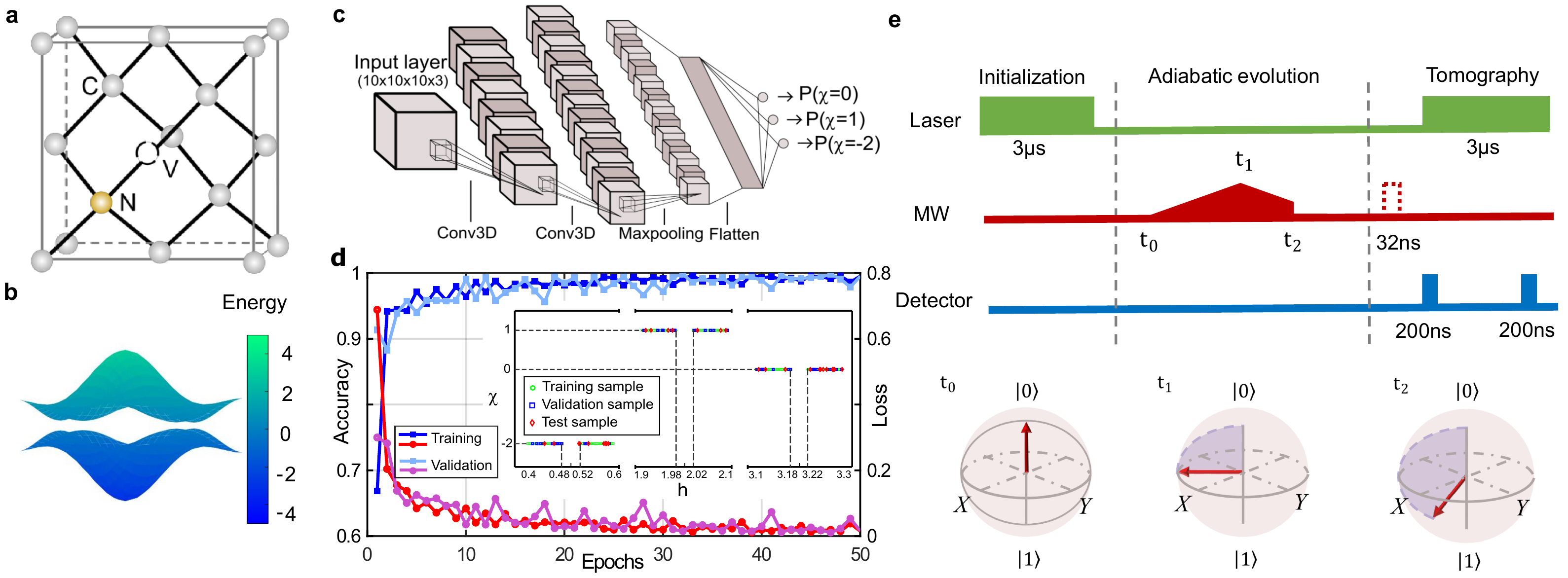}
 \caption{\textbf{Experimental setup and the topological phase classifier. }\textbf{a}  The structure of a nitrogen-vacancy center in a diamond crystal. \textbf{b} Theoretical energy dispersion of a two-band Hopf insulator for the momentum layer $k_z=\pi$ with $h=0.5$. Energies are calculated on a 20 $\times$ 20 grid in the $(k_x, k_y) $ plane. \textbf{c} The structure of the 3D convolutional neural network classifier. The input data are density matrices sampled from a $10\times 10\times 10$ regular grid in the momentum space; each density matrix is represented by three real indexes among the Bloch sphere. The classifier outputs the prediction confidences $P(\chi = 0,1,-2)$ for each phase. \textbf{d} The classifier is trained on numerically generated data from $h\in[-5,5]$ except intervals $[0.48, 0.52]$, $[1.98, 2.02]$ and $[3.18,3.22]$, where the experimental legitimate examples lie. There are $3471$ samples in the training set, $952$ samples in the validation set, and $521$ samples in the test set. The inset shows the distribution of the training, validation, and test samples. The accuracy and cross-entropy loss converge after about 10 epochs. \textbf{e} The experimental procedure for the preparation and measurement of the ground states of the Hopf Hamiltonian at each momentum $\mathbf k$. The dashed rectangle inserted before the final measurement represents the $\pi/2$ pulse with different phases. The directions of the electron spin on the Bloch sphere at three different time points are shown below the sequence. }
\label{fig: experiment setup}
\end{figure*}

However, the existence of adversarial examples \cite{Tygar2011Adversarial, Papernot2016Transferability, Biggio2018Wild, Goodfellow2015Explaining}, which are carefully crafted input samples that can deceive the learning model at a high confidence level, poses a serious concern about the reliability of machine-learning approaches as well. In particular, it has been shown recently that typical phase classifiers based on deep neural networks are indeed extremely vulnerable to adversarial perturbations---adding a tiny amount of carefully crafted noises into the original legitimate samples would mislead them to make incorrect predictions with a near-unit confidence \cite{jiang2019vulnerability}. Surprisingly, this is even true for learning topological phases of matter, where traditional wisdom would expect apparent robustness of the classifiers since topological invariants are immune to continuous deformations by definition. Yet, in spite of the antecedent theoretical progress, the experimental demonstration of adversarial examples in learning phases of matter, especially topological phases, is still lacking hitherto. In this work, we report such an experiment with a solid-state quantum simulator consisting of a single NV center in a diamond (see Fig. \ref{fig: experiment setup}). 

The NV center in diamond is a point defect \cite{doherty2013nitrogen}, consisting of a nitrogen atom that substitutes a carbon atom and a nearest-neighbor lattice vacancy, as shown in Fig. \ref{fig: experiment setup}\textbf{a}. It has long coherence time at room temperature and can be conveniently manipulated through lasers or microwaves, making this system versatile for applications in quantum networks \cite{Hensen2015Loopholefree,kalb2017entanglement,humphreys2018deterministic,pompili2021realization}, high-resolution sensing \cite{barry2020sensitivity,hsieh2019imaging,grinolds2013nanoscale,zaiser2016enhancing}, quantum information processing \cite{yao2012scalable,van2012decoherence,zhang2020efficient,bradley2019ten,wrachtrup2001quantum}, and quantum simulation \cite{georgescu2014quantum,choi2019probing,yuan2017observation}, etc. Here, with the NV center platform, we experimentally demonstrate adversarial examples in learning topological phases, with a focus on the peculiar Hopf insulators. More concretely, we first train a phase classifier based on deep convolutional neural networks (CNNs) so that it can classify different Hopf phases with an accuracy higher than $99.2\%$ on legitimate samples. We then show that, after adding a tiny amount of carefully-designed perturbations to the model Hamiltonian, this phase classifier would misclassify the experimentally generated adversarial examples with a confidence level up to $99.8\%$.  Through full tomography, we verify that the fidelity between the legitimate and corresponding adversarial samples is large (the average fidelity is $93.4\%$), ensuring that the adversarial perturbations added are small indeed. In addition, we also extract the topological invariant and topological links by traditional means and demonstrate that they are robust to adversarial perturbations.  This sharp robustness contrast between the traditional methods and machine-learning approaches clearly showcases the vulnerability aspect of the latter, highlighting the demand for in-depth investigations about the reliability of machine learning approaches in adversarial scenarios and for developing countermeasures.  

\vspace{.3cm}
\noindent\textbf{Results}

\begin{figure*}[htbp]
  \centering
  \includegraphics[width=1\textwidth]{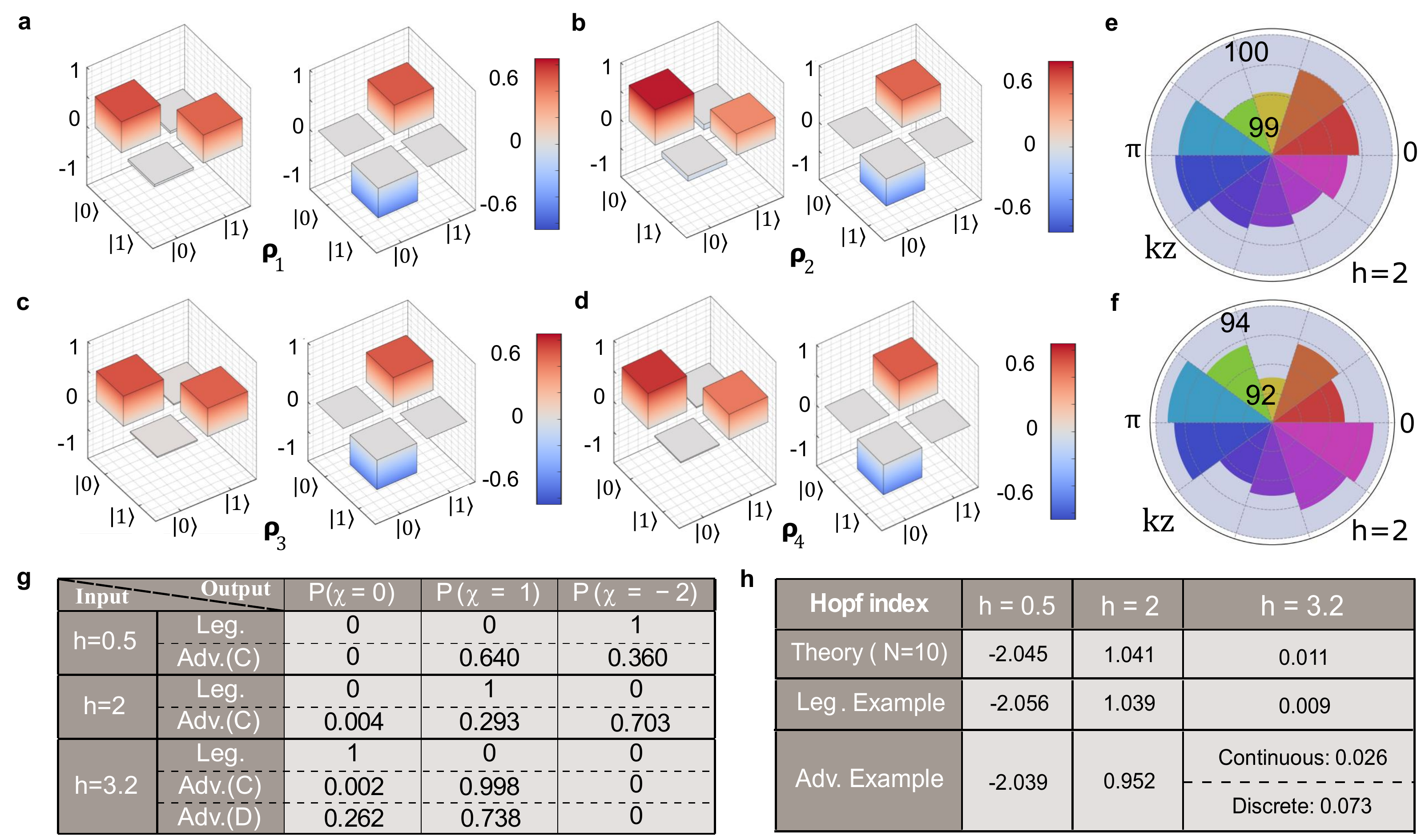}
  \caption{\textbf{The physical properties and performance of experimentally implemented adversarial examples.} \textbf{a} to \textbf{d} Density matrices of the spin states at $\mathbf{k}=(0.2\pi, 0.6\pi, 1.4\pi),h=0.5$.  \textbf{a}  Theoretical ground state of Hamiltonian $H_{\mathbf{k}}$, which is regarded as a single point in the legitimate sample. \textbf{b} The corresponding state  numerically generated by adding adversarial perturbations at the same point. \textbf{c} Experimental implementation of this legitimate sample at momentum $\mathbf{k}$. The fidelity $F(\rho_1, \rho_3)$ is $ 99.68(31)\%$.  \textbf{d} Experimental implementation of the adversarial example at momentum $\mathbf{k}$. The fidelity $F(\rho_2, \rho_4)$ is $99.23(26)\%$. \textbf{e} The average fidelity for each  $k_z$ layer in $[0:1.8\pi]$, $h=2$ between numerically generated and experimentally implemented adversarial examples. The angular direction represents the different layers of $k_z$ and the radial direction represents the fidelity, the average fidelity is $99.65(46)\%$. \textbf{f} The average fidelity for each $k_z$ layer in $[0:1.8\pi]$, $h=2$ between experimentally implemented legitimate samples and adversarial examples. The average fidelity is $93.40\%$. \textbf{g}  The classifier's output on experimentally implemented legitimate samples and their corresponding adversarial examples. The classifier successfully identifies legitimate examples' phases with nearly unity confidence but is fooled by experimentally implemented adversarial examples, incorrectly classifying their phases with confidence larger than $0.5$.
  \textbf{h} The Hopf index calculated by using the conventional discretization approach. As for discrete sampling on $10\times 10\times 10$ grids, the ideal cases are listed in the second row, where the small difference is caused by the discretization error of the 3D momentum integration. The Hopf index for both experimentally implemented legitimate and adversarial examples is close to its ideal integer values for the corresponding topological phases.}
 \label{fig: experimental results}
\end{figure*}

\noindent\textbf{Machine learning of topological phases.}  To experimentally implement adversarial examples and demonstrate the vulnerability of machine learning in topological phases, we first train a phase classifier based on deep neural networks  to predict topological phases with high accuracy. Concretely, we focus on an intriguing three-dimensional (3D) topological insulator called the Hopf insulator \cite{Moore2008Topological, Deng2013Hopf}, whose Hamiltonian in the momentum space reads:
\begin{equation}  
 H_{{\rm TI}}=\sum_{\mathbf{k}\in {\rm BZ}}\Psi_{\mathbf{k}}^{\dagger}H_{\mathbf{k}}\Psi_{\mathbf{k}}=\sum_{\mathbf{k}}\Psi_{\mathbf{k}}^{\dagger} \mathbf{u_{k}}\cdot{\bm{\sigma}}\Psi_{\mathbf{k}},
\label{eq: Hamiltonian}
\end{equation}  
where $\mathbf{u_{k}}=(u_{x},u_{y},u_{z})$ with $u_{x} =2(\sin k_{x}\sin k_{z}+C_{\mathbf{k}} \sin k_{y})$,
$u_{y}=2(C_{\mathbf{k}}\sin k_{x}-\sin k_{y}\sin k_{z})$, and $u_{z}=\sin^{2}k_{x}+\sin^{2}k_{y}-\sin^{2}k_{z}-C_{\mathbf{k}}^{2}$; $ \Psi_{\mathbf{k}}^{\dagger}=(a_{\mathbf{k},\uparrow}^{\dagger},a_{\mathbf{k},\downarrow}^{\dagger})$ are fermionic annihilation operators with pesudo-spin states $\mid\uparrow \rangle $ and $\mid\downarrow \rangle$ at each momentum $\mathbf{k}$ in the Brillouin zone (BZ); $\bm{\sigma}=(\sigma_{x},\sigma_{y},\sigma_{z})$ are Pauli matrices, $C_{\mathbf{k}}\equiv\cos k_{x}+\cos k_{y}+\cos k_{z}+h$. The energy dispersion with $h=0.5$ is shown in Fig. \ref{fig: experiment setup}\textbf{b}. Hopf insulators are peculiar 3D topological insulators that originate from the mathematical theory of Hopf fibration and elude the standard periodic table for topological insulators and superconductors for free fermions \cite{Kitaev2009Periodic,Schnyder2008Classification}. They can manifest the deep connection between knot theory and topological phases of matter in a visualizable fashion \cite{yuan2017observation,Deng2018Probe}. Their topological properties are characterized by a topological invariant (the Hopf index) $\chi$ \cite{Moore2008Topological, Deng2013Hopf} and direct calculations show that $\chi=-2$ if $| h |<1$, $\chi=1$ if $1<| h | <3$, and $\chi=0$ otherwise. 

We numerically generate $5000$ samples with varying $h\in[-5,5]$ and train a 3D CNN in a supervised fashion. Figure \ref{fig: experiment setup}\textbf{c} shows the structure of CNN classifier we use. The input data are reconstructed density matrices in the momentum space and outputs are classification confidences $P(\chi = 0,1,-2)$ of each possible phase. The training process is shown in Fig. \ref{fig: experiment setup}\textbf{d} and the inset shows the distribution of the training, validation, and test samples. After training, the classifier obtains near-perfect performance on the numerically generated data, with accuracies of $99.2\%$ and $99.6\%$ on the validation and training datasets, respectively.  

\vspace{.2cm}
\noindent\textbf{Experimental implementation.} We use a single NV center as a simulator to experimentally implement the model Hamiltonian in Eq. (\ref{eq: Hamiltonian}). The ground state of the NV electron spin consists of $| m_s=0\rangle$ and degenerate $| m_s=\pm 1\rangle$ state with zero-field splitting of $2.87$ GHz \cite{childress2006coherent,bar2013solid}. Our setup is based on a home-built confocal microscope with an oil-immersed lens. To enhance photon collection efficiency, a solid immersion lens is fabricated on top of the NV center. A magnetic field of 472 Gauss is applied along the NV symmetry axis to polarize the nearby nuclear spins and remove degeneracy between states $| m_s=\pm 1\rangle$. We use the subspace $| m_s=0\rangle$ and $| m_s=-1\rangle$, denoted by $| 0\rangle$, $| -1\rangle$ of the electron spin in the experiment. The experiment sequence is shown in Fig. \ref{fig: experiment setup}\textbf{e}. First, the spin state is initialized to $|0\rangle$ by optical pumping. Then, a microwave (MW) is applied to adiabatically evolve the spin state. By tuning the amplitude, frequency, and phase of MW,  the electron spin is evolved to the ground state of the corresponding Hamiltonian at a given momentum point $\mathbf{k}$ \cite{xu2017experimental,lian2019machine}. The electron spin states  at three different evolution time points are shown
below the sequence in Fig. \ref{fig: experiment setup}\textbf{e}. After the adiabatic evolution, quantum state tomography is performed, and the state density matrices are retrieved via maximum likelihood estimation \cite{james64measurement}.

To obtain  density matrices sampling of the Hamiltonian in Eq.(\ref{eq: Hamiltonian}), we mesh the momentum space $\mathbf{k}=(k_x, k_y, k_z)$ into $10 \times 10 \times 10$ grids with equal spacing. We use ground state density matrices of $H_{\mathbf{k}}$  with  $h=0.5,2,3.2$ as legitimate samples. These legitimate samples are used as ground truth to evaluate the topological phase classifier and latter generate corresponding adversarial examples. Partial of our results are plotted in Fig. \ref{fig: experimental results}. For clarity, we show the legitimate sample calculated by Eq. (\ref{eq: Hamiltonian}) in Fig. \ref{fig: experimental results}\textbf{a} and corresponding experimental implementation in Fig. \ref{fig: experimental results}\textbf{c} at the point $\mathbf{k}=({0.2\pi,0.6\pi,1.4\pi}), h=0.5$. To alleviate the effect of experimental Gaussian noise on adversarial perturbations, we implement all reconstructed states at very high fidelity: For all three legitimate samples with $h=3.2, 2, 0.5$, the average fidelities are $99.77(41)\%$, $99.78(41)\%$, and $99.77(45)\%$, respectively (for the fidelity on each momentum point, see Fig. S2 in the Supplementary Materials). The table in  Fig. \ref{fig: experimental results}\textbf{g} presents the classifier's output for three legitimate samples, which are all correct classifications with nearly unity confidence. 

\begin{figure*}[htbp]
 \centering
 \includegraphics[width=16cm]{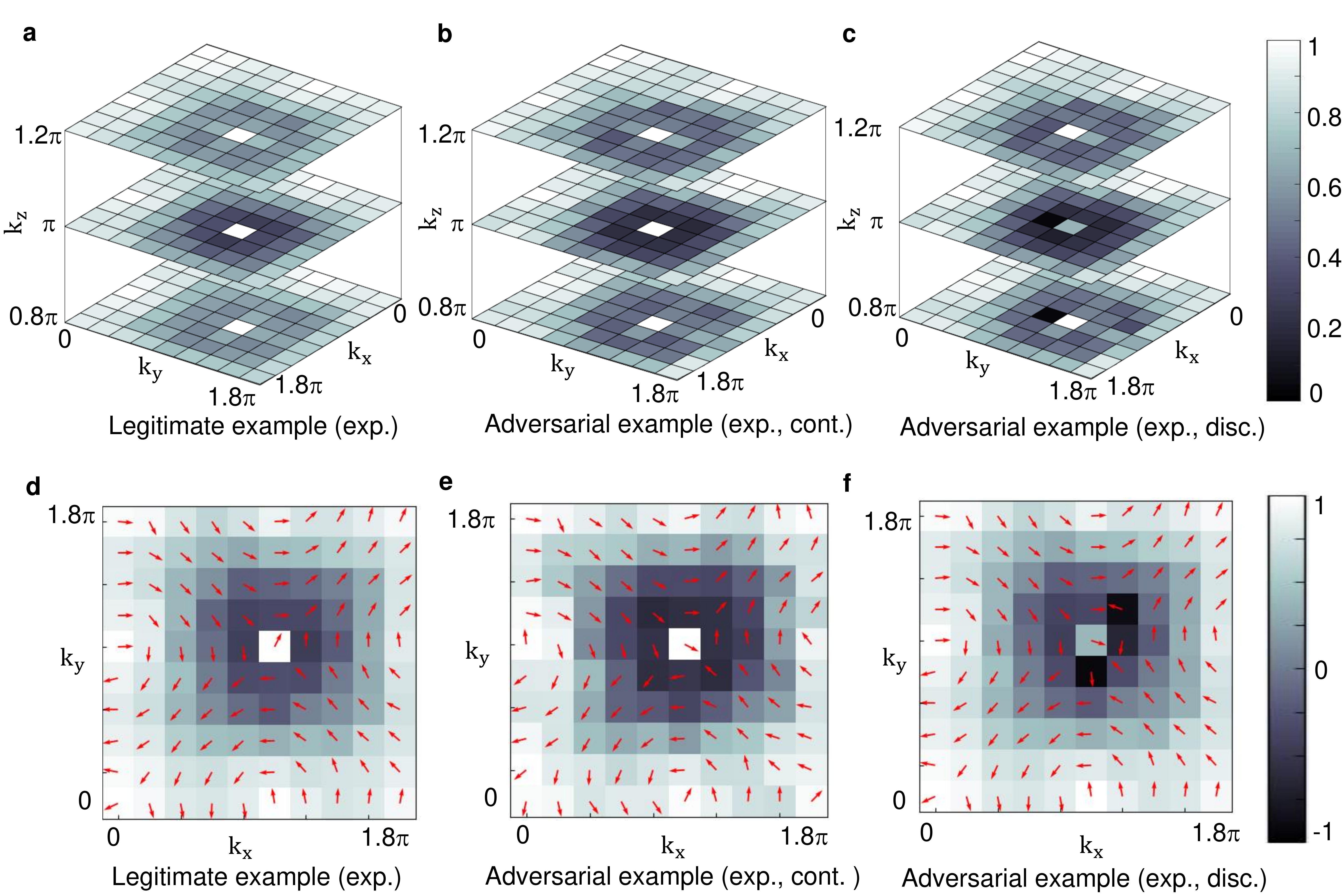}
 \caption{\textbf{Visualization of  experimentally realized density matrices for the Hopf insulators with $\mathbf{h=3.2}$. } \textbf{a} to \textbf{c} The first component's magnitude of the input data for $h=3.2$ at $k_{z}=0.8\pi$, $\pi$ and $1.2 \pi$. \textbf{a} Legitimate sample with $h=3.2$ implemented in the experiment. \textbf{b} Adversarial examples realized in the experiment with continuous perturbations generated by the projected gradient descent method. The average fidelity for the experimentally implemented legitimate samples is over $98\%$ with fidelity distribution shown in the Supplementary Materials. \textbf{c} Adversarial examples realized in the experiment with discrete perturbation generated by the differential evolution algorithm. Among 1000 density matrices as input, only seven of them have been changed and successfully mislead the classifier. \textbf{d} to \textbf{f} Measured spin texture for $k_z=\pi$, $h=3.2$. For each subfigure, $k_x$ and $k_y$ vary from 0 to $1.8\pi$ with equal spacing of $0.2\pi$. At each momentum $\textbf{k}$, the state can be represented on the Bloch sphere. The arrows in the plane show the direction of the Bloch vector projected to the $x-y$ plane. The colors label the $z$ component of the Bloch vector. \textbf{d} Legitimate sample with $h=3.2$ implemented in experiment. \textbf{e} Adversarial examples implemented in the experiment with continuous perturbations generated by the momentum iterative method. \textbf{f} Adversarial examples implemented in  the experiment with discrete perturbation generated by the differential evolution algorithm.
 }
 \label{fig: show slices}
\end{figure*}

The existence of adversarial examples may rise severe problems for machine-learning approaches to the classification of different phases of matter.  Choosing the loss function $\boldsymbol{L}$ as the metric to evaluate the performance of the classifier, we first search for numerical adversarial examples which can mislead the classifier by solving an optimization problem \cite{jiang2019vulnerability}: finding a bounded perturbation $\delta$ adding to the legitimate samples' data to maximize the loss function $\boldsymbol{L}$. We employ various strategies to approximately solve this optimization problem, including the fast gradient sign method (FGSM) \cite{Madry2018Towards}, projected gradient descent (PGD) \cite{Madry2018Towards}, momentum iterative method (MIM) \cite{Dong2018Boosting} and differential evolution algorithm (DEA) \cite{Storn1997DE, Das2011DE,Su2019One}. Concretely, we apply PGD and MIM to obtain continuous adversarial perturbations based on all three legitimate samples \cite{papernot2016cleverhans}, and apply DEA to obtain one adversarial example with discrete perturbations based on the legitimate sample with $h=3.2$ (see the Supplementary Materials S6 for more details). 

We note that the existence of numerical adversarial examples does not guarantee that we can implement them in real experiments due to inevitable experimental imperfections---the experimental noises may wash out the bounded and carefully-crafted adversarial perturbations. In fact, as shown in Refs. \cite{Jeremy2019Certified, Li2019Certified} certain noises that are typical in experiments would nullify the adversarial examples. 
In the worst case, when the noise is in the opposite direction of the adversarial perturbation, the experimentally implemented adversarial examples will no longer be able  to mislead the classifier. To this end, we numerically simulate the experimental noises acting on these numerically obtained adversarial examples and examine their performances on the classifier. After the simulation, for each scenario we select one example with the strongest robustness against experimental noise and reconstruct its corresponding Hamiltonian. Illustrating densities for the numerical and experimentally implemented adversarial examples are shown in Fig. \ref{fig: experimental results}\textbf{b} and \textbf{d}. Figure \ref{fig: experimental results}\textbf{e} shows the average fidelity for each $k_z$ layer with $h=2$. All adversarial examples have high fidelities (larger than $93\%$) but the classifier incorrectly predicts their phases with high confidence level, as shown in Fig. \ref{fig: experimental results}\textbf{g}. \\

\noindent\textbf{Demonstration of adversarial examples.} In the previous section, we illustrate that the experimentally implemented adversarial examples, which have high fidelity to original legitimate data, can mislead the  topological phase classifier. In this section, we further demonstrate the effectiveness of these adversarial examples from the physical perspective. In Fig. \ref{fig: show slices}\textbf{a}-\textbf{c}, with $h=3.2$, we visualize experimentally implemented density matrices of legitimate samples, adversarial examples with continuous perturbations and adversarial examples with discrete perturbations. From the comparison, the obtained adversarial examples look almost the same as the original legitimate ones. It is surprising that even local discrete changes, as shown in Fig. \ref{fig: show slices}\textbf{c}, can mislead the classifier to make incorrect prediction. This result is at variance with the physical intuition that Hopf insulators are robust to local perturbations due to their topological nature \cite{Moore2008Topological, Deng2013Hopf}, indicating that the neural network based classifier does not fully captured the underlying topological characteristics \cite{jiang2019vulnerability}.

Focusing on classifying topological phases of Hopf insulators, we expect that the adversarial perturbations should not change the topological properties, including the integer-valued topological invariant and the topological links associated with the Hopf fibration \cite{Moore2008Topological}. We use a conventional method, which is based on the experimentally measured data, to probe the Hopf index \cite{Fukui2005Chern, Deng2014Direct}. The results are shown in Fig. \ref{fig: experimental results}\textbf{h} (see Sec. S4 of the Supplementary Materials for more details). We find that each adversarial example's Hopf index has only a negligible  difference to the corresponding legitimate ones, all close to the correct integer numbers. 

\begin{figure}[]                                                         
 \centering
 \includegraphics[width=8.5cm]{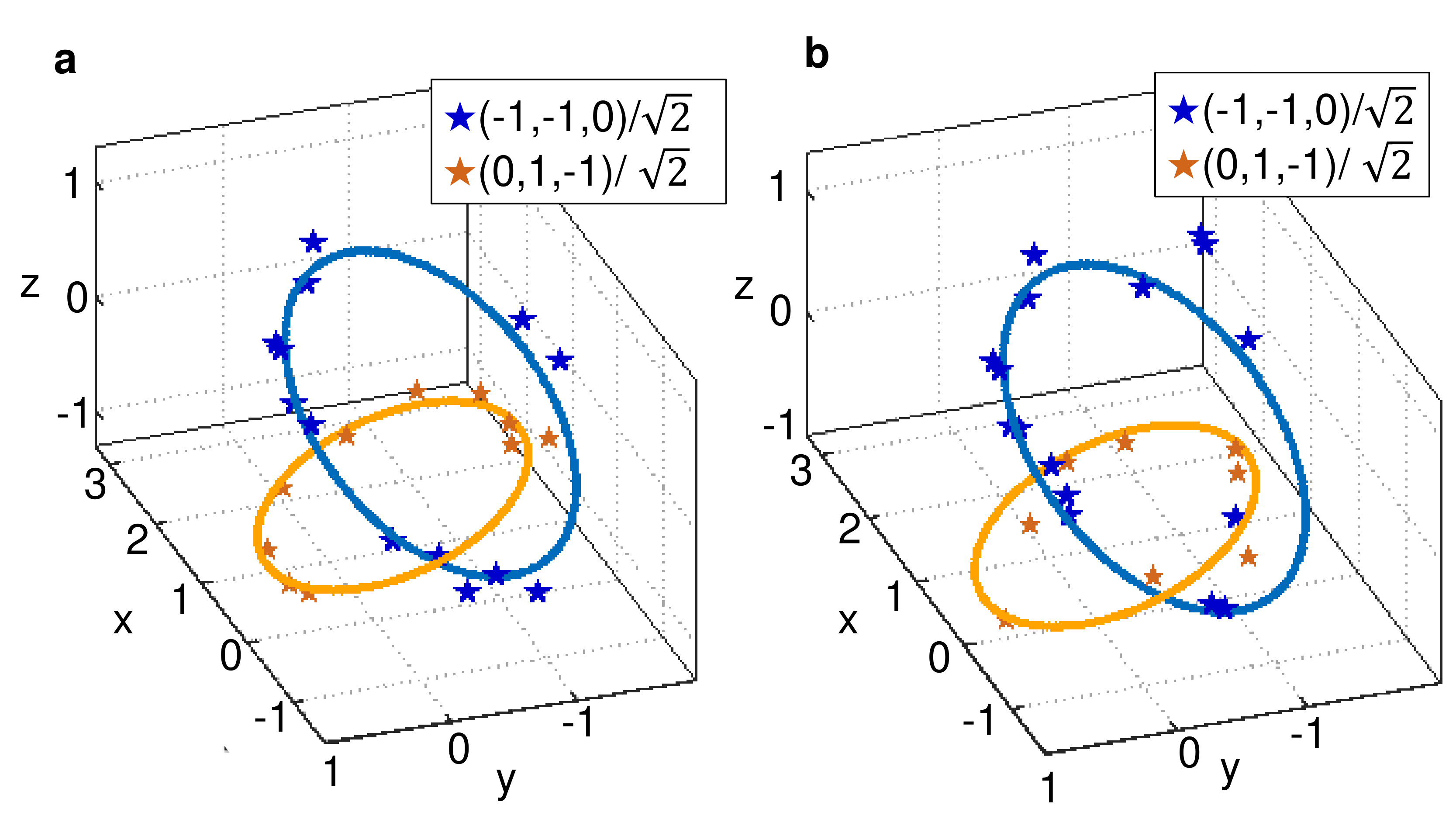}
 \caption{\textbf{The 3D preimage contours which show topological links for the Hopf insulator with $\mathbf{h=0.5}$.} \textbf{a} Topological link, obtained from two spin states with Bloch sphere representations $\mathbf{S}=(-1,-1,0)/\sqrt{2}$ (blue) and $(0,1,-1)/\sqrt{2}$ (orange), for a legitimate sample in the stereographic coordinates of $\mathbb{S}^3$, between two spin states on the Bloch sphere. Solid lines are curves from numerically calculated directions $\mathbf{S}_{\text{th}}$ and the stars are experimentally measured spin orientations $\mathbf{S}_{\text{exp}}$. The deviation $|\mathbf{S}_{\text{exp}}-\mathbf{S}_{\text{th}}|\leq0.25~(0.35)$ for the blue (orange) curve. \textbf{b} The topological link for an adversarial example with deviation $|\mathbf{S}_{\text{exp}}-\mathbf{S}_{\text{th}}|\leq0.25$ for blue and orange curves. }
 \label{fig: show contour} 
\end{figure}

The momentum-space spin texture of Hopf insulators harbors a knotted structure, which is called the Hopfion \cite{faddeev1997stable}. In Fig. \ref{fig: show slices}\textbf{d}-\textbf{f}, we show cross sections of the measured spin textures before and after adding continuous and discrete adversarial perturbations. The spin textures present an illustration on 3D twisting of the Hopfion, which keep original structure and are almost not affected by adversarial perturbations. A more intuitive demonstration of Hopf links can be derived if we consider the preimage of a fixed spin orientation on the Bloch sphere, which will form a closed loop in the momentum space $\mathbb{T}^3$. For topological nontrivial phases, the loops for different orientations are always linked \cite{yuan2017observation}. As shown in Fig. \ref{fig: show contour}, we visualize the 3D preimage contours of legitimate and adversarial examples in $\mathbb{R}^3$ from the stereographic coordinates of $\mathbb{S}^3$, with $h=0.5$, orientations $\mathbf{S}=(-1,-1,0)/\sqrt{2}$ and $(0,1,-1)/\sqrt{2}$ on the Bloch sphere (see Sec. S7 of the Supplementary Materials for more details). We observe that the loops are correctly linked together as $h=0.5$ corresponds to topological nontrivial $\chi=-2$ phase, for both legitimate and adversarial examples. This result illustrates that the adversarial perturbations do not affect the Hopf link, despite the fact that they alter the predictions of the classifiers drastically. 

\vspace{.3cm}
\noindent \textbf{Discussion}

\noindent The above sections clearly showcase that the adversarial perturbations do not affect the physical properties of topological phases. The incorrect predictions given by the classifier indicate that the classifier does not learn the accurate and robust physical criterion for identifying topological phases. which is consistent with the theoretical prediction in the recent paper \cite{jiang2019vulnerability}.  
How to exploit the experimental data to make the classifier better learn physical principles, and how to protect the neural-network based classifiers in real experiments from adversarial perturbations, are interesting and important problems worth further investigation. 
In addition, recent experiments have demonstrated the simulation of non-Hermitian topological phases with the NV platform \cite{zhang2020observation} and a theoretical work on learning non-Hermitian phases with exotic skin effect in an unsupervised fashion has also been reported \cite{Yu2021Unsupervised}. In the future, it would be interesting and desirable to study adversarial examples for unsupervised learning of topological phases, both in theory and in experiment.   

In summary, we have experimentally demonstrated the  adversarial examples in learning topological phases with a solid-state simulator. Our result showcases that neural network-based classifiers are vulnerable to tiny and carefully-crafted perturbations, even with the protection introduced by experimental noises. We implemented the adversarial examples in experiment and studied their properties: high fidelity, unchanged topological invariant, and topological link. These results reveal that current machine learning methods do not fully capture the underlying physical principles and thus are especially vulnerable to adversarial perturbations. 

\vspace{.3cm}
\noindent\textbf{{Methods}}

\small{\noindent\textbf{Experimental setup.} Our experiment is implemented on a home-built confocal microscope at room temperature. The 532 nm diode laser passes through an acoustic optical modulator setting in a double-pass configuration. The laser can be switched on and off on the time scale of $\sim$ 20 ns with on-off ratio to 10000:1. A permanent magnetic  provides the static magnetic field of 472 Gauss, the magnetic field is precisely aligned parallel to the symmetry axis of the NV center by observing the emitted photon numbers \cite{epstein2005anisotropic}. With this magnetic field, a level anticrossing in the electronic state allows electron-nuclear spin flip-flops, which polarizes the nuclear spin \cite{Jacques2009Dynamic}.  The magnetic field also removes the degeneracy between $| m_s=\pm 1\rangle$ states. The spin state is first initialized by $3~\mu s$ laser excitation, then MW modulation is implemented by programming two orthogonal 100 MHz carrier signals, which are generated by arbitrary waveform generator. MW is amplified and guided through coaxial cables to gold coplanar waveguide close to the NV center. The emitted photons are collected through an oil lens-immersion objective lens (NA=1.49) and detected by an avalance photodiode. Spin state is readout by counting the spin-dependent number of photons. To enhance collection efficiency, a solid immersion lens with $6.74~\mu m$ diameter is fabricated. The fluorescence count is about 260 kcps under 0.25 mW laser excitation, with the signal-noise ratio is about ${\rm 100:1}$. The sequence is repeated $7.5\times 10^5$ times, collecting about $3.9\times10^4$ photons.
\\ \hspace*{\fill} \\
\noindent\textbf{Adiabatic passage approach.} Consider the electron subspace of $|0\rangle$ and $|-1\rangle$ state, in a rotating frame, the effective Hamiltonian with variable time $t$ is
\begin{equation}
H_{\rm{eff}}=
\begin{pmatrix}
0&|\Omega| e^{i\varphi}\\
|\Omega| e^{-i\varphi}& -\Delta\omega(t)
\end{pmatrix},
\end{equation}
$\Omega(t)$ is the MW amplitude, $\varphi$ is the MW phase, $\Delta\omega(t)=\omega_{0}-\omega_{\rm MW}$, and $\omega_{0}$ and $\omega_{\rm MW}$ are NV resonant frequency and MW frequency, respectively. 
The Hamiltonian $H_{\rm{TI}}$ is similarly expressed as:

\begin{equation}
H_{\rm TI}=
\begin{pmatrix}
0&u_x-iu_y\\
u_x+iu_y & -u_z
\end{pmatrix}.
\end{equation}
We terminate the adiabatic evolution at time $t_{c}$ to satisfy $\Delta\omega(t_{c})/\Omega(t_{c})=u_{z}/\sqrt{u_x^2+u_y^2}$. Phase $\varphi = -{\rm arctan}(u_{y}/u_{x})$ is kept constant in the adiabatic evolution. To satisfy the adiabatic condition \cite{lian2019machine}:
\begin{equation}
    \left\vert \frac{\hbar\langle\psi^{(e)}|\dot{\psi}^{(g)}\rangle}{E_{e}-E_{g}} \right\vert\ll1,
\end{equation}
where $|\psi^{g}\rangle$ and $|\psi^{e}\rangle$ are ground and excited states of the Hamiltonian $H_{\rm TI}$, $E_{g}$ and $E_{e}$ are energies of the ground state and excited state. We use $\Omega_{\rm max}=2\pi\times7.81$ MHz and $\Delta\omega_{\rm max}=2\pi\times10$ MHz during the adiabatic passage process for a total  time of $1500$ ns.}
\\ \hspace*{\fill} \\
\noindent\textbf{Data availability}

\noindent \small{All data needed to evaluate the conclusions in the paper are present in the paper and/or the Supplementary Materials. Additional data related to this paper may be requested from the corresponding authors.}

\noindent\textbf{Acknowledgements} 
\noindent \small{This work was supported by the Frontier Science Center for Quantum Information of the Ministry of Education of China, Tsinghua University Initiative Scientific Research Program, and the Beijing Academy of Quantum Information Sciences. D.-L. D. also acknowledges additional support from the Shanghai Qi Zhi Institute. }
\\ \hspace*{\fill} \\
\noindent\textbf{Author Contributions} 
\noindent \small{H.Z. carried out the experiment under the supervision of L.-M. D. and D.-L.D.. S.J. did the numerical simulations and analyzed the experimental data together with H.Z. All authors contributed to the experimental set-up, the discussions of the results and the writing of the manuscript. }  
\section{reference}

\clearpage
\setcounter{figure}{0}
\makeatletter
\renewcommand{\algorithmicrequire}{\textbf{Input }}  
\renewcommand{\algorithmicensure}{\textbf{Output}}  
\newtheorem{corollary}{Corollary} \newtheorem{definition}{Definition} %
\newtheorem{example}{Example} 
\newtheorem{lemma}{Lemma} %
\newtheorem{proposition}{Proposition} \newtheorem{theorem}{Theorem} %
\newtheorem{fact}{Fact} \newtheorem{property}{Property}
\renewcommand{\thefigure}{S\@arabic\c@figure}
\setcounter{equation}{0} \makeatletter
\renewcommand \theequation{S\@arabic\c@equation}
\renewcommand \thetable{S\@arabic\c@table}
\renewcommand \thealgorithm{S\@arabic\c@algorithm}
\begin{center} 
{\large \bf Supplementary Materials for: Experimental demonstration of adversarial examples in learning topological phases}
\end{center} 

\section{Section S1. Experimental Setup}
This experiment is implemented on a home-built confocal microscope at room temperature. The diamond sample (type IIa, Element 6) is grown via chemical vapor deposition and cut along $\langle 100\rangle$ orientation, with the natural abundance of $1.1\%$ ${\rm^{13}C}$.  The 532 nm diode laser (Coherent Compass 315M) passes through an acoustic optical modulator (AOM, Isomet 1250C-848) setting in a double-pass configuration. This configuration can increase the on-off ratio to 10000:1 and constrain the leakage of the laser. Then the laser is reflected by a diachronic mirror (DM) into an oil-immersed objective lens (NA=1.49, Olympus),  and the focused spot size is about $300 \times 300$ $nm^2$ on the diamond. The lens is mounted on a three-axis closed-loop piezo (Physik Instrumente,  E-725) with a scanning range of $100 \times 100 \times 100$ $\mu m^3$. The photons in the wavelength ranging from 637 to 800 nm pass through the same objective lens and DM, then are collected by a single-mode fiber and detected by a single photon counting module  (SPCM, Excelitas, SPCM-AQRH-14-FC).  A permanent magnetic provides the static magnetic field of 472 Gauss, by observing the emitted photon numbers, we can align the magnetic field parallel to the NV symmetry axis.  The magnitic field removes the degeneracy between $|m_s=\pm 1\rangle$ states and polarizes the nuclear spin with the polarization rate typically exceeding 95$\%$ \cite{epstein2005anisotropic}.

The microwave (MW) is generated by an analog signal generator (Agilent, N5181B) and modulated by an IQ mixer (Marki IQ 1545LMP) with two orthogonal 100 MHz carrier signals, which are generated from the analog output of an arbitrary waveform generator (AWG,Tektronic 5014C, sample rate 1 GHz). The combined signal then is amplified by a high power amplifier (Mini Circuits, ZHL-16w-43-S+) and delivered into a gold coplanar waveguide close to the NV center. The amplitude of the combined signal is in the linear range of the amplifier to ensure that the Rabi frequency is proportional to the amplitude of the combined signal.  

In our experiment, a single sequence starts with  3 $\mu$s laser excitation for the electron spin polarization and ends with 3 $\mu$s laser pulse for the spin state detection. The signal photons are collected for 200 ns right after the detection laser rises and reference photons are collected for 200 ns after 2 $\mu$s. The sequences are programmed and loaded to the AWG. To enhance collection efficiency, a solid immersion lens with 6.74 $\mu m$ diameter is fabricated by a focused ion beam (FEI company, Helios nanolab 660). The 
photon number is about 260 kcps under 0.25 mW laser excitation, increasing the signal-noise ratio to about ${\rm 100:1}$.  The sequence is repeated $7.5\times 10^5$ times, collecting about $3.9\times10^4$ photons.

\section{Section S2. Adiabatic Passage}
Consider the electron subspace of $|0\rangle$ and $|-1\rangle$ states, in a rotating frame, the effective Hamiltonian with variable time $t$ is
\begin{equation}
H(t)=\Omega(t)(S_{x}\cos\varphi - S_{y}\sin \varphi)+\Delta\omega(t) S_{z},
\label{NVHamltonian}
\end{equation}
where $\Omega(t)$ is the MW amplitude, $\varphi$ is the MW phase, and $\Delta\omega(t)=\omega_{0}-\omega_{\rm MW}$, with $\omega_{0}$ and $\omega_{\rm MW}$ being the resonant frequency and MW frequency, respectively; $S_{x,y,z}$ are given as the following (setting $\hbar=1$):
\begin{equation}
S_{x}=\frac{1}{2}
\begin{pmatrix}
0&1\\
1&0
\end{pmatrix}
,S_{y}=\frac{1}{2}
\begin{pmatrix}
0&-i\\
i&0
\end{pmatrix}
,S_{z}=
\begin{pmatrix}
0&0\\
0&-1
\end{pmatrix}
.
\end{equation}
We apply  the adiabatic passage process described as
\begin{eqnarray}
\Omega(t)&=&
\begin{cases}
2\Omega_{\rm max}t/T,  & t \le T/2,\\
2\Omega_{\rm max}(1-t/T) &t>T/2,
\end{cases}
\\
\Delta\omega(t)&=&\Delta\omega_{\rm max} - 2 \Delta\omega_{\rm max}\ t/T.
\end{eqnarray}
The frequency and amplitude are shown in Supplementary Figure\ref{microwave}.

\begin{figure}[htp]
\centering
 \includegraphics[width=5.2cm]{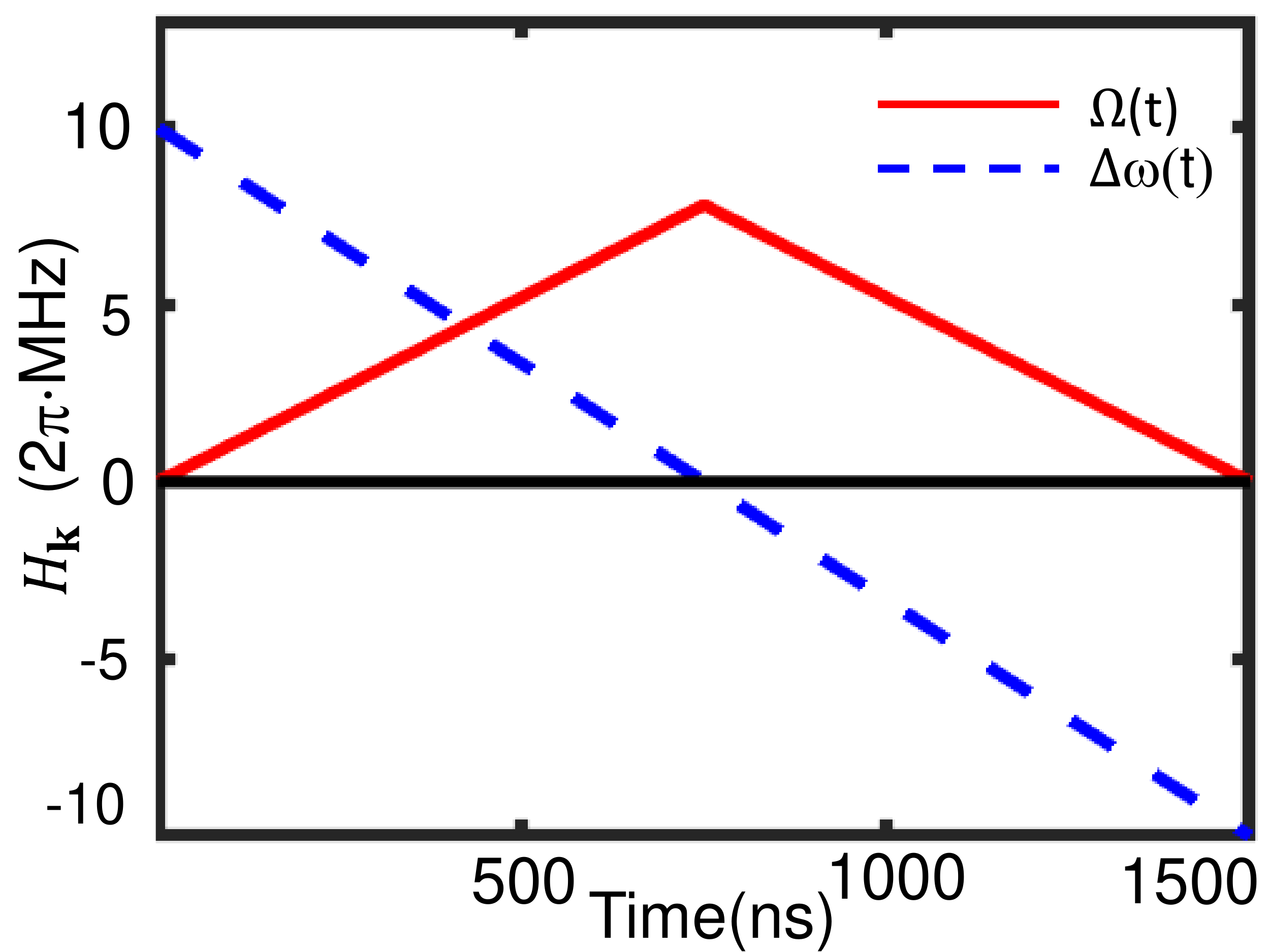}
 \caption{\textbf{The MW frequency and amplitude in the adiabatic passage.}}
\label{microwave}
\end{figure}
Comparing with the Hamiltonian in Eq. (\ref{NVHamltonian}), we terminate the adiabatic passage process at time $t_{c}$ to satisfy $\Delta\omega(t_{c})/\Omega(t_{c})=u_{z}/\sqrt{u_x^2+u_y^2}$. Phase $\varphi = -{\rm arctan}(u_{y}/u_{x})$ is kept constant during  the adiabatic passage process. To satisfy the adiabatic condition \cite{lian2019machine}:
\begin{equation}
Q\equiv\frac{2(\Delta\omega(t)^{2}+|\Omega_2(t)|^2)^{3/2}}{||\dot{\Omega}(t)|\Delta\omega(t)-|\Omega(t)|\dot{\Delta\omega}(t)|}\gg1,
\end{equation}
in our experiment, we use $\Omega_{\rm max}=2\pi\times7.81$ MHz and $\Delta\omega_{\rm max}=2\pi\times10$ MHz. During the adiabatic passage process, we keep $Q>25$  in our experiment. 

\section{Section S3. Quantum State Tomography}
\begin{figure}[htp]
  \centering
  \includegraphics[width=8.5cm]{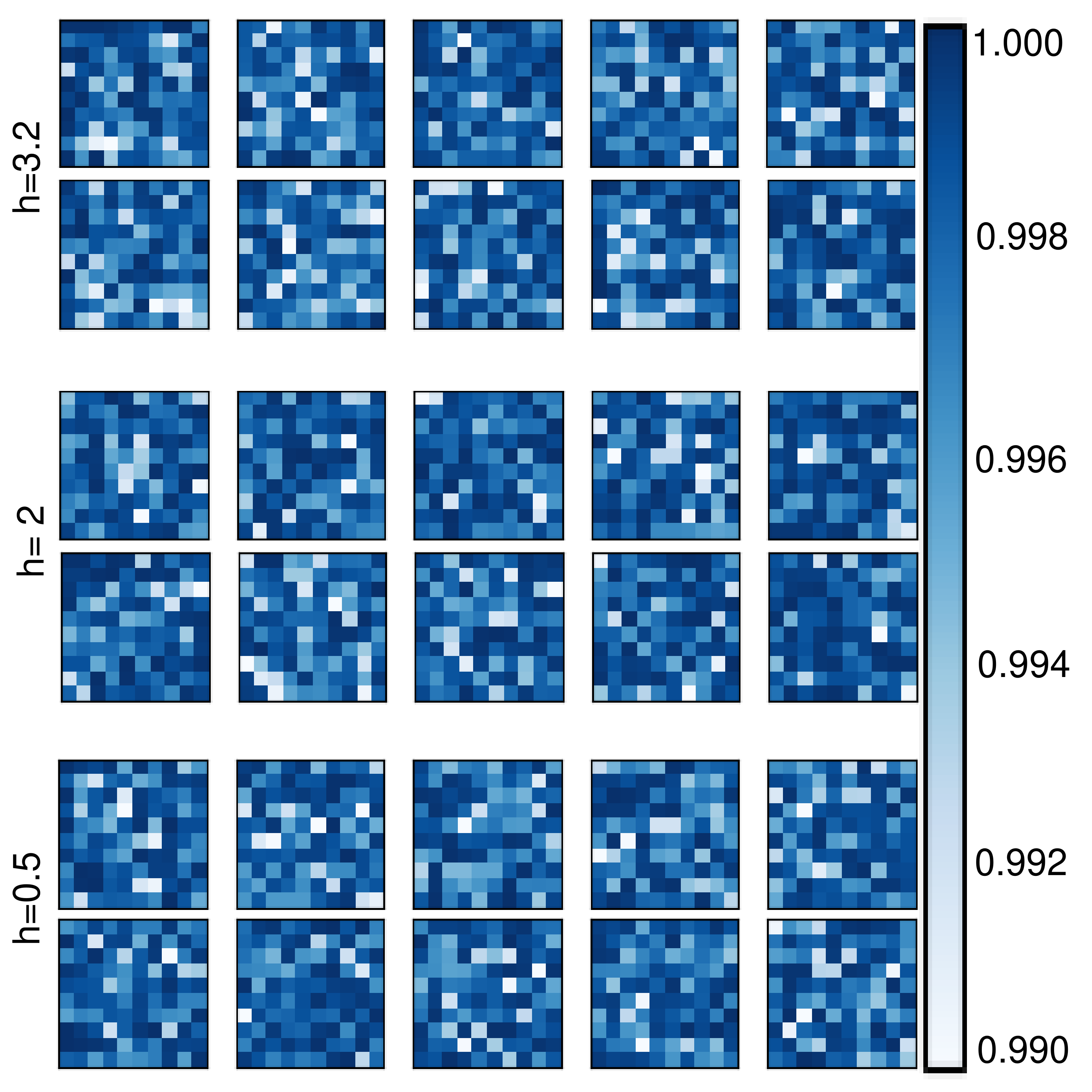}
\caption{\textbf{State fidelity $F_{\mathbf{k}}$ for legitimate samples.} The states are measured via the likelihood estimation at different momenta $\mathbf{k}$. Each panel consists of 10 subfigures where the momenta $k_{z}$ are equally spaced from 0 to $1.8\pi$. The horizontal and vertical axes of each subfigure denote  $k_{x}$ and $k_{y}$ varying from 0 to $1.8\pi$ with equal spacing. The panels represent parameters $h=3.2$ with average fidelity 99.77(41) $\%$, $h=$2 with average fidelity 99.78(41) $\%$ and $h=0.5$ with average fidelity 99.77(45) $\%$, respectively.}
\label{FidelityLegitimateAll}
\end{figure}

\begin{figure}[htp]
  \centering
  \includegraphics[width=8.5cm]{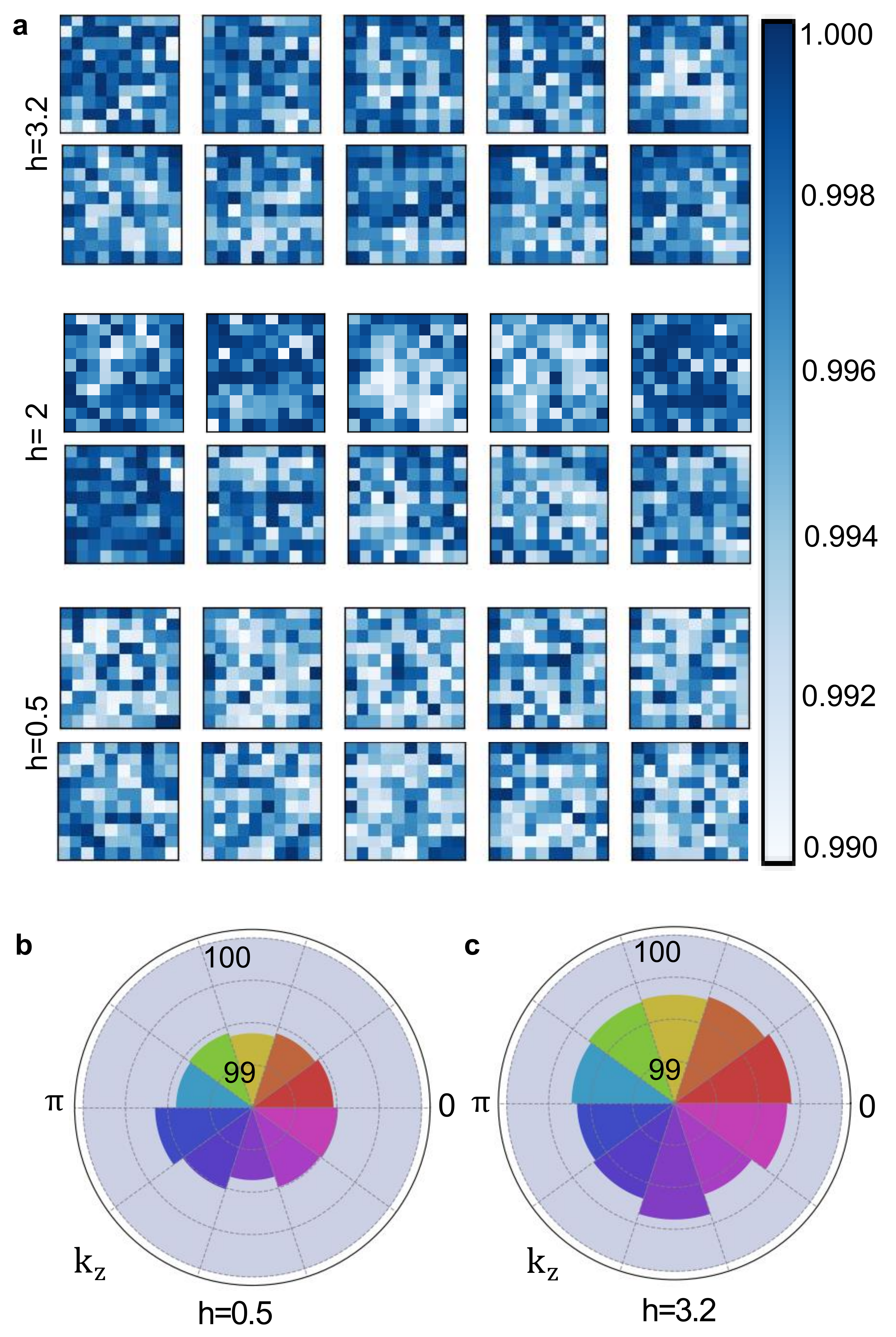}
\caption{\textbf{State fidelity $F_{\mathbf{k}}$ for adversarial examples.} \textbf{a} The states are measured via the likelihood estimation  at different momenta $\mathbf{k}$. Each panel plots 10 subfigures with the momenta $k_{z}$ equally spaced from 0 to $1.8\pi$. The horizontal and vertical axes of each subfigure denote  $k_{x}$ and $k_{y}$ varying from 0 to $1.8\pi$ with equal spacing. The panels represent parameters $h=3.2$ with average fidelity 99.64(43) $\%$, $h=$2 with average fidelity 99.65(46) $\%$ and $h=0.5$ with average fidelity 99.48(46) $\%$, respectively. \textbf{b} The average fidelities at different $\mathbf{k_z}$ layers for $h=0.5$. The angular direction represents different $k_z$ and the radius direction represents the fidelity. \textbf{c} The average fidelities at different $\mathbf{k_z}$ layers for and $h=3.2$.}.
\label{FidelityadversaryAll}
\end{figure}

We perform a full quantum state tomography to measure the eigenstates of the topological Hamiltonian at each momentum $\mathbf{k}$. The photon numbers in the basis $\{\rm Z\}$ is directly measured and in bases $\{\rm{+X, +Y, -X}\}$, a $\pi/2$ pulse with phase $\{\rm{0,-\pi/2,\pi}\}$ will be inserted before the detection.
In a single-qubit system, to ensure the non-negative definite, Hermitian, and trace-one properties of a density matrix, the density matrix $\hat{\rho}$ can be written  as
$\hat{\rho}_{t}=T^{\dagger}(t)T(t)/{\rm Tr}\{T^{\dagger}(t)T(t)\}$, where $T$ reads:
\begin{equation}
T(t)=
\begin{pmatrix}
t_1&0\\
t_2+it_3&t_4
\end{pmatrix}.
\end{equation}
The measurement consists of a set of four coincidence photon numbers. In our experiment the expected photon number for the $\nu$-th measurement is
\begin{equation}
\bar{n}_{\nu}(t_1,t_2,t_3,t_4)=N_{0}p_{0,\nu}+N_{-1}(1-p_{0,\nu}),
\end{equation}
where $N_{0}$ and $N_{-1}$ are the photon numbers of states $|0\rangle$ and $|-1\rangle$ measured by fast Rabi oscillation, and $p_{0}$ denotes the probability for the spin at $|0\rangle$.  Assuming the noise on this coincidence measurements has a Gaussian probability distribution, the probability $P(n_1, n_2, n_3, n_4)$ of photon numbers $\{n_1, n_2, n_3, n_4\}$ produced by the density matrix $\hat{\rho}_{p}(t_1,t_2,t_3,t_4)$ reads:
\begin{equation}
P(n_1, n_2, n_3, n_4)=\prod_{\nu=1}^{4}{\rm exp}\bigg[-\frac{(n_{\nu}-\bar{n}_{\nu})^2}{2\sigma_{\nu}^2}\bigg],
\end{equation}
where $\sigma_{\nu}$ is the standard deviation of photon numbers for the $\nu$-th measurement (approximated to be $\sqrt{\bar{n}_{\nu}}$) and ${n}_{\nu}$ is photon numbers measured in different tomography bases.
The optimization problem reduces to finding the minimum of the following function \cite{james64measurement}:
\begin{equation}
\mathcal{L}(t_1, t_2, t_3, t_4)=\sum_{\nu=1}^{4}\frac{[\bar{n}_{\nu}(t_{1}, t_{2}, t_{3}, t_{4})-n_{\nu}]^2}{2\bar{n}_{\nu}(t_1, t_2, t_3, t_4)}.
\end{equation}
 Error bars are calculated through Monte Carlo simulations by assuming a Poissonian distribution of the photon numbers. Supplementary Figure \ref{FidelityLegitimateAll} and Supplementary Figure \ref{FidelityadversaryAll}\textbf{a} show the average fidelity for different Hopf index for legitimate samples and adversarial examples, respectively. For legitimate samples, the  average fidelity for $h=3.2$, $h=2$ and $h=0.5 $ are  $99.77(41)~\%$, 99.78(41) $\%$, and 99.77(45) $\%$, respectively. Similarly, for adversarial examples, the average fidelity for $h=3.2$, $h=2$ and $h=0.5$ are 99.64(43) $\%$, 99.65(46) $\%$, and 99.48(46) $\%$, respectively. We also show the average fidelity for each $k_z$ layer with $h=0.5$ and $h=3.2$ between the numerically generated and experimentally implemented states in Supplementary Figure \ref{FidelityadversaryAll}\textbf{b} - \textbf{c}.

\section{Section S4. Discretization Scheme to Measure Hopf Index}
We use the discretization scheme introduced in \cite{Fukui2005Chern, Deng2014Direct} and applied in \cite{yuan2017observation} to measure the Hopf index directly from experimental data. The Hopf index can be written as:
\begin{equation}
\chi=-\int_{\text{BZ}}\bm{F}\cdot \bm{A}d^3\mathbf{k},
\end{equation}
where $\bm{F}$ is the Berry curvature with $F_\mu = \frac{i}{2\pi}\epsilon_{\mu\nu\tau}(\partial_{k_\nu}\langle\psi_{\mathbf{k}}|)(\partial_{k_\tau}|\psi_{\mathbf{k}}\rangle)$ and $\bm{A}$ is the associated Berry connection satisfying $\nabla\times \bm{A}=\bm{F}$. To avoid the arbitrary phase problem, we can use a discretized version of the Berry curvature \cite{Fukui2005Chern, Deng2014Direct}:
\begin{equation}
    F_{\mu}(\mathbf{k}_{\bm{J}}) = \frac{i}{2\pi}\epsilon_{\mu\nu\tau}\ln{U_{\nu}(\mathbf{k}_{\bm{J}}})\ln{U_{\tau}(\mathbf{k}_{\bm{J}+\hat{\nu}}}).
\end{equation}
The $U(1)$-link is defined as
\begin{equation}
U_{\nu}(\mathbf{k}_{\bm{J}})=\frac{\langle\psi(\mathbf{k}_{\bm{J}})|\psi(\mathbf{k}_{\bm{J}+\hat{\nu}})\rangle}{|\langle\psi(\mathbf{k}_{\bm{J}})|\psi(\mathbf{k}_{\bm{J}+\hat{\nu}})\rangle|},
\end{equation}
with $\bm{\hat{\nu}}\in\{\bm{\hat{x}},\bm{\hat{y}},\bm{\hat{z}}\}$, which is a unit vector in the corresponding direction. This discretized version of $\bm{F}$ can be calculated after performing quantum state tomography at all points $\mathbf{k}_{\bm{J}}$ on the momentum  grid. We can also obtain the Berry connection $\bm{A}$ by Fourier transforming $\nabla\times\bm{A}=\bm{F}$ with the Coulomb gauge $\nabla\cdot{\bm{A}}=0$. Finally, instead of doing the integral, we sum over all points on the momentum grid to obtain the Hopf index $\chi$. It is shown in  \cite{yuan2017observation,lian2019machine} that for a $10\times10\times10$ grid, this method is quite robust to various perturbations and can extract the Hopf index with high accuracy.

\section{Section S5. Convolutional Neural Network Classifier}
We use a 3D convolutional neural network (CNN) classifier to predict the Hopf index. The classifier accepts density matrices on a 10$\times$10$\times$10 momentum grid as input. Each density matrix is represented by three real indexes $(x_1,x_2,x_3)$ in the Bloch sphere, where
\begin{equation}
    x_i=\text{tr}(\rho\sigma_i),\quad i=1,2,3,
\end{equation}
with $\sigma_i\; (i=1,2,3)$ denoting the usual Pauli matrices.  With the activation function set as the Relu function, the classifier consists of an input layer with the shape 10$\times$10$\times$10$\times$3, two 3D convolution layers, one max pooling layer, and one fully-connected flattening layer. The classifier ends with a softmax layer, which outputs the classification confidences $P(\chi = 0, 1, -2)$ for each topological phase. The detail of parameters used in the classifier are listed in the Supplementary Table \ref{table: classifier structure}.

We adapt the supervised learning approach to training our classifier learning topological phases \cite{Carrasquilla2017Machine, vanNieuwenburg2017Learning,Chng2017Machine}.  With $h$ uniformly varied from $-5$ to $5$, we numerically generate 5000 samples with known Hopf index $\chi$. To avoid the numerically generated data being too close to experimental legitimate samples with $h=0.5,2,3.2$, we remove numerical samples in the intervals $[0.48,0.52]$, $[1.98,2.02]$, and $[3.18,3.22]$. We randomly choose samples to form a training set with size 3471, a validation set with size 952, and a test set with size 521. The hyper-parameters used in the training process are shown in the Supplementary Table \ref{table: Hyper-parameters}.

\begin{table}[htp]
\center
\begin{tabular}{cccc}
\toprule
Layer & Size/Ch. & Stride & Act. \\
\midrule
Input & 3 & &\\
Conv3D $2\times2\times2$ & 8 & 1 & Relu \\
Conv3D $4\times4\times4$ & 16 & 1 & Relu \\
MaxPool $2\times2\times2$& & 1 & \\
Flatten & & & \\
Linear & 512 & & Relu \\
Linear & 3 & & Softmax \\
\bottomrule
\end{tabular}
\caption{\textbf{The architecture details of the classifier.} Size represents the number of the hidden nodes used in fully-connected layers. Ch. represents the number of channels used in convolutional layers. Act. represents the activation function.}
\label{table: classifier structure}
\end{table}

\begin{table}[htp]
\center
\begin{tabular}{c|c|c|c|c}
\toprule
Optimizer & Epoch & Batch size & Learning rate & $\rho$ (in RMSprop)\\
\midrule
RMSprop & 50 & 128 & $10^{-3}$ & 0.9\\
\bottomrule
\end{tabular}
\caption{\textbf{The hyper-parameters used for training the classifier.}}
\label{table: Hyper-parameters}
\end{table}

The training process ends with an accuracy of $99.2~\%$ on the training set and $99.6~\%$ on the validation set. On the test set, the classifier obtains an accuracy of $99.2~\%$ and the outputs for each sample are shown in Supplementary Figure \ref{test_Pr}. One can identify the correct phase transition points $h=-3,-1,1,3$ from the sharp probabilities crosses.
\begin{figure*}[htp]
 \centering
 \includegraphics[width=16cm]{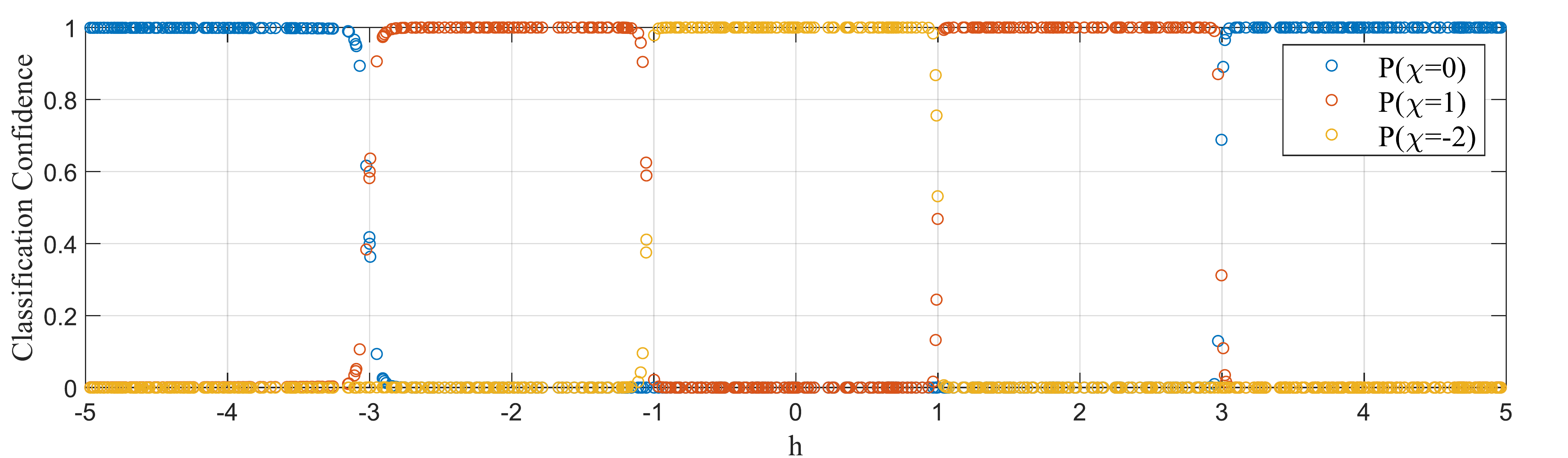}
 \caption{\textbf{The classifier's output on the test set.} For each input as the density matrices on the $10\times10\times10$ momentum grid, the classifier outputs the probabilities for each phase that the input may belong to. The classifier exhibits near-perfect performance on the data which it has not seen before, and successfully identify the transition points by the sharp confidence clips around $|h|=3$ and $1$.}
 \label{test_Pr}
\end{figure*}

\section{Section S6. Adversarial Examples Generation}

In the main text, we claim that our CNN classifier is vulnerable to adversarial examples, which are obtained by adding a tiny but carefully-crafted perturbation on the legitimate data. In the supervised learning case, as the legitimate samples are labeled by $(\textbf{x}_{\text{leg}}, y_{\text{leg}})$, the task to generate adversarial examples reduces to maximize the loss function on legitimate samples:
\begin{equation}
    \max_{\bm{\delta}\in\Delta}L(f(\bm{x}_{\text{leg}}+\bm{\delta};\theta),y_{\text{leg}}).
\end{equation}
In this work, we apply three typical methods in the adversarial machine learning literature to solve this problem: projected gradient descent (PGD) \cite{Madry2018Towards}
and momentum iterative method (MIM) \cite{Dong2018Boosting} for the continuous attack scenario; differential evolution algorithm (DEA) \cite{Storn1997DE, Das2011DE} for the discrete attack scenario.

\subsection{Projected gradient descent}
Projected gradient descent (PGD) is developed based on an elementary method called the fast gradient sign method (FGSM) \cite{Madry2018Towards}. The FGSM is a one-step attack that perturbs the legitimate sample $\bm{x}_{\text{leg}}$ according to the sign of the gradient:
\begin{equation}
\delta_{\text{FGSM}}=\epsilon\cdot\text{sign}(\nabla_x L(f(\bm{x}_{\text{leg}};\theta),y_{\text{leg}})).
\end{equation}

With a large step size $\epsilon$, the FGSM may perform poorly at the point whose  gradient changes abruptly. To deal with this problem, PGD uses FGSM with a multiple-step procedure. In each step, PGD performs a projection $\pi_C$ to ensure that the perturbation is restricted in a certain region \cite{Madry2018Towards}:
\begin{equation}
bm{x}^{(t+1)}_{\text{leg}}=\pi_C(\bm{x}^{(t)}_{\text{leg}}+\frac{\epsilon}{T}\cdot\delta_{\text{FGSM}}(\bm{x}^{(t)}_{\text{leg}})), \quad t=1,2\dots, T.
\end{equation}

In our work, we restrict the perturbation to be within the region with $l_{\infty}$-norm $\gamma$, which means that for each component $x_j$ of $\bm{x_{\text{leg}}}$, $\pi_C$ projects it into $[x_{\text{leg},j}-\gamma, x_{\text{leg},j}+\gamma]$. The pseudo-code for PGD is shown in Algorithm \ref{PGD_pseudocode}.

\begin{algorithm}[htp]  
\caption{Projected Gradient Descent Method}
\label{PGD_pseudocode}
\begin{algorithmic}[1]  

\Require The classifier $f(\bullet;\theta)$, loss function $L$, the legitimate density matrix $(\bm{x_{\text{leg}}},y)$.
\Require The FGSM step size $\epsilon$, iteration number $T$, $l_{\infty}$-norm restriction $\gamma$.
\Ensure An adversarial example $\bm{x^*}$.
\State $\bm{x^{(0)}}=\bm{x_{\text{leg}}}$
\State  $\alpha=\frac{\epsilon}{T}$
\For {$i=1,\dots,T$}
	\State $g^{(i)}=\nabla_xL(f(\bm{x^{(i-1)}};\theta),y_{\text{leg}}))$
	\For {Each component $j$ of $\bm{x^{(i-1)}}$}
		\State $\delta_j=\alpha\cdot \text{sign}(g^{(i)}_j)$
		\State $x^{(i)}_j=x^{(i-1)}_j+\delta_j$
		\If {$x^{(i)}_j>x^{(0)}_j+\gamma$}
			\State $x^{(i)}_j=\pi_C(x^{(i)}_j)=2(x^{(0)}_j+\gamma)-x^{(i)}_j$
		\EndIf
		\If {$x^{(i)}_j<x^{(0)}_j-\gamma$}
			\State $x^{(i)}_j=\pi_C(x^{(i)}_j)=2(x^{(0)}_j-\gamma)-x^{(i)}_j$
		\EndIf
	\EndFor
\EndFor \\
\Return $\bm{x^*}=\frac{\bm{x}^{(T)}}{||\bm{x}^{(T)}||}$
\end{algorithmic}
\end{algorithm}

\subsection{Momentum iterative method}
PGD performs FGSM iteratively with a much smaller step size to deal with rapidly changing gradients, but it can easily drop into local extremums and exhibit poor performance. The momentum iterative method (MIM) introduces momentum into the iterative FGSM to avoid being trapped by local extremums. Concretely, with $l_{\infty}$-norm $\epsilon$, iteration number $T$, the MIM updates the adversarial example with the following rule \cite{Dong2018Boosting}:
\begin{equation}
a_0 = 0, \quad
a_t = \mu\cdot a_{t-1} + \frac{\nabla_xL(f(\bm{x^{(t-1)}};\theta),y_{\text{leg}})}{||\nabla_xL(f(\bm{x^{(t-1)}};\theta),y_{\text{leg}})||},
\end{equation}
\begin{equation}
\bm{x^{(t)}}=\bm{x^{(t-1)}}+\frac{\epsilon}{T}\cdot\text{sign}(a_t),
\end{equation}
where $\mu$ is a decay factor and $a$ performs as the accelerated velocity which contains the information of past gradient descent direction. The pseudo-code for MIM is shown in Algorithm \ref{MIM_pseudocode}.

\begin{algorithm}[htp]  
\caption{Momentum Iterative Method}
\begin{algorithmic}[1]  

\Require The classifier $f(\bullet;\theta)$, loss function $L$, legitimate density matrix $(\bm{x_{\text{leg}}},y)$.
\Require The $l_{\infty}$-norm restriction $\epsilon$, iteration number $T$, decay factor $\mu$.
\Ensure An adversarial example $\bm{x^*}$.
\State $\bm{x^{(0)}}=\bm{x_{\text{leg}}}$
\State  $\alpha=\frac{\epsilon}{T}$
\State $a_0=0$
\For {$i=1,\dots,T$}
	\State $a_i=\mu\cdot a_{i-1} + \frac{\nabla_xL(f(\bm{x^{(i-1)}};\theta),y_{\text{leg}})}{||\nabla_xL(f(\bm{x^{(i-1)}};\theta),y_{\text{leg}}))||}$
	\State $\bm{x^{(t)}}=\bm{x^{(t-1)}}+\alpha\cdot\text{sign}(a_t)$
\EndFor \\
\Return $\bm{x^*}=\frac{\bm{x}^{(T)}}{||\bm{x}^{(T)}||}$
\end{algorithmic}  
\label{MIM_pseudocode}
\end{algorithm}

\subsection{Differential evolution algorithm}

Differential evolution algorithm (DEA) belongs to
the general class of evolutionary algorithms, which is powerful to  solve complex multi-modal optimization problems. Concretely, DEA first randomly generates $n$ candidates $\{X\}=X_1,X_2,\dots,X_n$ as possible solutions. These candidates become parents in the first iteration. In each iteration, DEA generates a new set of candidates called children from the current parents first by \cite{Storn1997DE, Das2011DE}:
\begin{equation}
X_i'=X_j+F\cdot(X_k-X_l),
\end{equation}
where $X_j,X_k,X_l$ are randomly picked from $\{X\}$ and distinct from each other. $F$ is called the mutual factor. Then, for each component $j$ of $X_i'$, with randomly picked $r_j$ from $U(0,1)$:
\begin{equation}
(X_i')_j=(X_i)_j,\quad \text{if } r_j>P,
\end{equation}
where $P$ is called the crossover probability. If the children $X_i'$ has better performance than the parent $X_i$, then DEA will replace $X_i$ with $X_i'$. This procedure will repeat for several iterations until the candidates are almost converged.

In our scenario, we need to discretely change $m$ density matrices on the $10\times10\times10$ grid and keep others unchanged. Therefore, the candidates are in the form of
\begin{equation}
\begin{aligned}
X&=(\textbf{pos}_1, \textbf{idx}_1)\times(\textbf{pos}_2, \textbf{idx}_2)\times\dots\times(\textbf{pos}_m, \textbf{idx}_m)\\&=(\textbf{pos},\textbf{idx})^m
\end{aligned},
\end{equation}
which changes $\bm{x_{\text{leg}}}$ by setting $(\bm{x_{\text{leg}}})_{\textbf{pos}_i}=\textbf{idx}_i$. The pseudo-code for DEA is shown in Algorithm \ref{DEA_pesudo}.

\begin{figure}[htp]
  \centering
  \includegraphics[width=8.5cm]{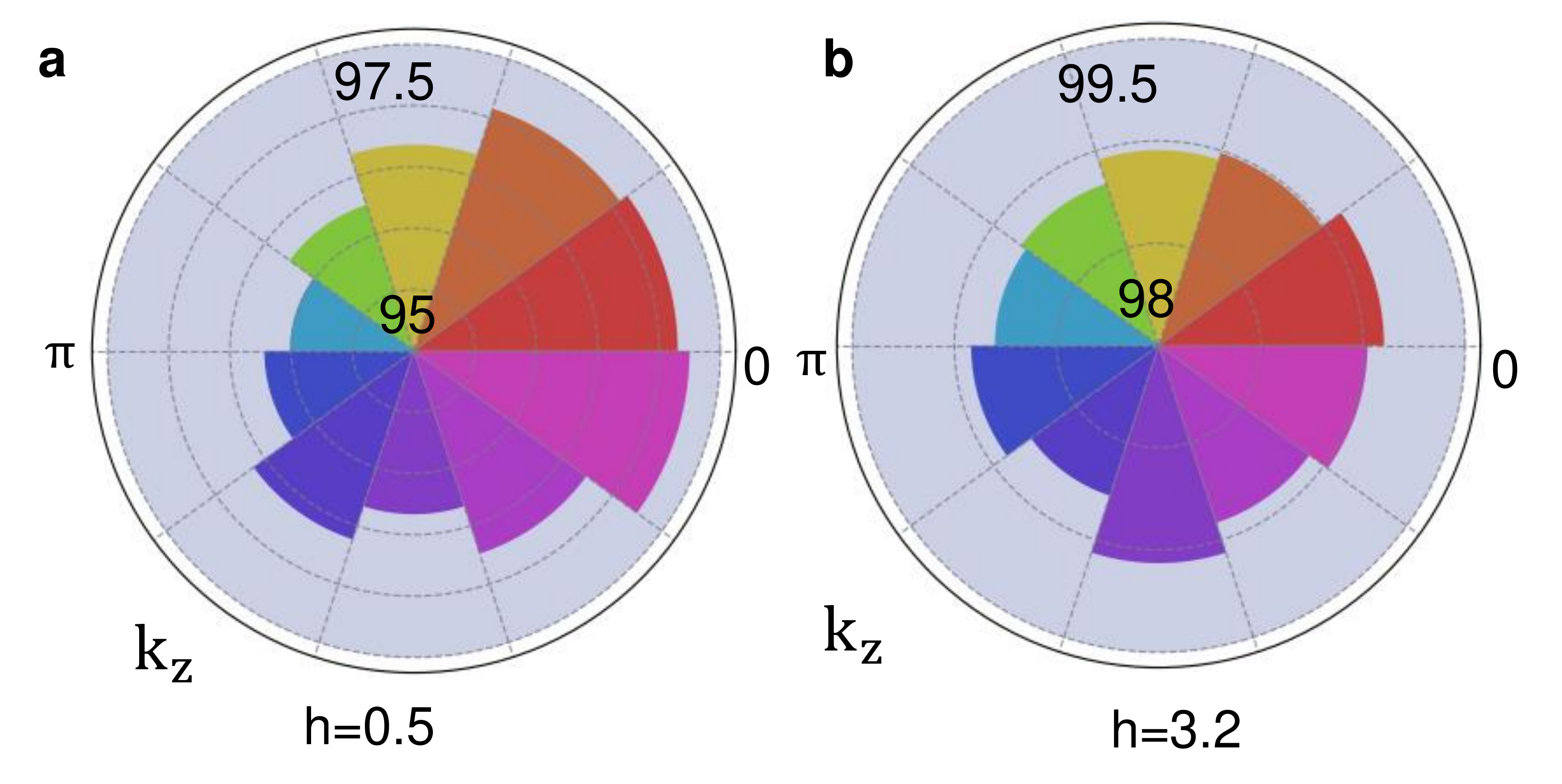}
\caption{\textbf{State fidelity $F_{\mathbf{k}}$ between experimentally implemented adversarial examples and legitimate samples.} \textbf{a} The fidelity at different $\mathbf{k_z}$ layers with $h=0.5$. The angular direction represents different $k_z$ and the radius direction represents the fidelity. \textbf{b} The fidelity at different $\mathbf{k_z}$ layers with $h=3.2$. }
\label{barsimilarity}
\end{figure}

\vspace{0.5cm}
We use cleverhans \cite{papernot2016cleverhans} to implement PGD and MIM, and adapt the code in Ref. \cite{Su2019One} to implement DEA. We use MIM for legitimate samples with $h=0.5,2$ and use PGD for $h=3.2$ to numerically generate adversarial examples with continuous perturbations on all density matrices. We also use DEA for $h=3.2$ to numerically generate adversarial examples with only seven discrete density matrices changed. It should be noted that, although all perturbations generated by PGD, MIM, and DEA are restricted to be tiny, they can make the legitimate samples unphysical, namely that they are not normalized to one after adding the perturbation. To avoid this, we renormalize all density matrices in the adversarial examples, but this can make these examples lose the ability to mislead the classifier. The noises in experiment can also deteriorate their performance. To deal with this problem, we repeatedly adjust the parameters and restrictions to numerically generate different adversarial examples until there is one can successfully mislead the classifier after normalization and adding simulated noises. After numerically generating these adversarial examples, we reconstruct their Hamiltonian in the NV center. The fidelity between experimentally realized adversarial examples and legitimate samples at different $k_z$ layers are shown in Supplementary Figure \ref{barsimilarity}. The average fidelity for $h=0.5$ and $h=3.2$ are $96.63~\%$ and $98.94~\%$.

\begin{algorithm}[htp]
\caption{Differential Evolution Algorithm}
\begin{algorithmic}[1]  
\Require The legitimate sample $(\bm{x_{\text{leg}}},y)$, trained model $f(\bullet;\theta)$.
\Require The iteration number $T$, the population size $n$, the number $m$ of pixels to be changed, the mutual factor $F$, the crossover probability $P$.
\Ensure An adversarial example $\bm{x^*}$.
\State Randomly generate perturbation $X_i=(\textbf{pos},\textbf{idx})^m$ with $\textbf{pos}\in [0,9]^3$ and $\textbf{idx}\in [-1, 1]^3$ for $i=1,2,\dots,n$.
\State $\text{len}=6m=\text{lengnth of each }X_i$
\State Perform $X_i$ on $\bm{x_{\text{leg}}}$ to obtain $\bm{x_i^{*}}$ for $i=1,2,\dots,n$.
\For {$t=1,2,\dots,T$}
    \For {$i=1,2,\dots,n$}
        \State Randomly pick distinct $j,k,l\in [n]/\{i\}$
        \State $X_i'=X_j+F\cdot(X_k-X_l)$
        \For {$s=1,2,\dots,\text{len}$}
            \State Randomly pick $r$ from $U(0,1)$
            \If {$r>P$}
                \State $(X_i')_s=(X_i)_s$
            \EndIf
        \EndFor
        \State Perform $X_i'$ on $\bm{x_{leg}}$ to obtain $\bm{x_i^{*'}}$
        \If {$L(f(\bm{x_i^{*'}};\theta),y_{\text{leg}})>L(f(\bm{x_i^{*}};\theta),y_{\text{leg}})$}
        \State $X_i=X_i'$
        \State $x_i^*=x_i^{*'}$
        \EndIf
    \EndFor
\EndFor
\State Find the $x_q^*$ that has the largest $L(f(\bm{x_i^{*'}};\theta),y_{\text{leg}})$ among $\{x_1^*,x_2^*,\dots,x_n^*\}$\\
\Return $\bm{x^*}=\frac{x_q^*}{||x_q^*||}$
\end{algorithmic}  
\label{DEA_pesudo}
\end{algorithm}

\begin{figure*}[htp]
 \centering
 \includegraphics[width=15cm]{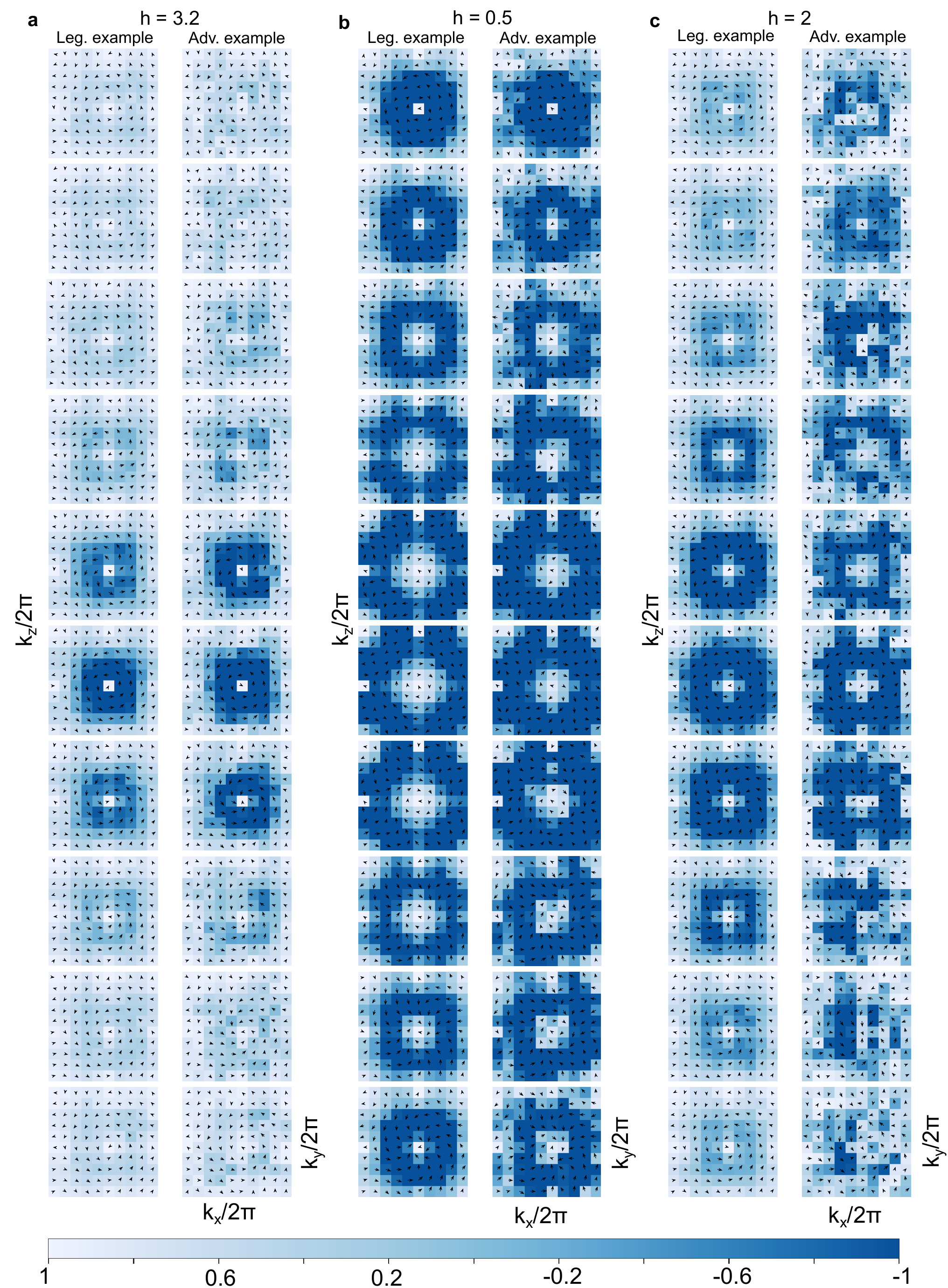}
 \caption{\textbf{Measured spin textures of experimental legitimate samples and  adversarial examples obtained by continuous attacks.} Each column contains all layers of measured spin textures for $k_z=0,0.1,\dots,0.9\times 2\pi$. The arrow at each point shows the projection of the Bloch vector into the x-y plane and the color represents the magnitude of the z component of the Bloch vector. \textbf{a} Topologically trivial phase with $h=3.2$, corresponding to $\chi=0$. \textbf{b} Topological nontrivial phase with $h=0.5$, corresponding to $\chi=-2$. \textbf{c} Topological nontrivial phase with $h=2$,   corresponding to $\chi=1$.}
 \label{allSlice}
\end{figure*}

\section{Section S7. 3D Hopf Fibration Representation}

\begin{figure}[htp]
 \centering
 \includegraphics[width=8.5cm]{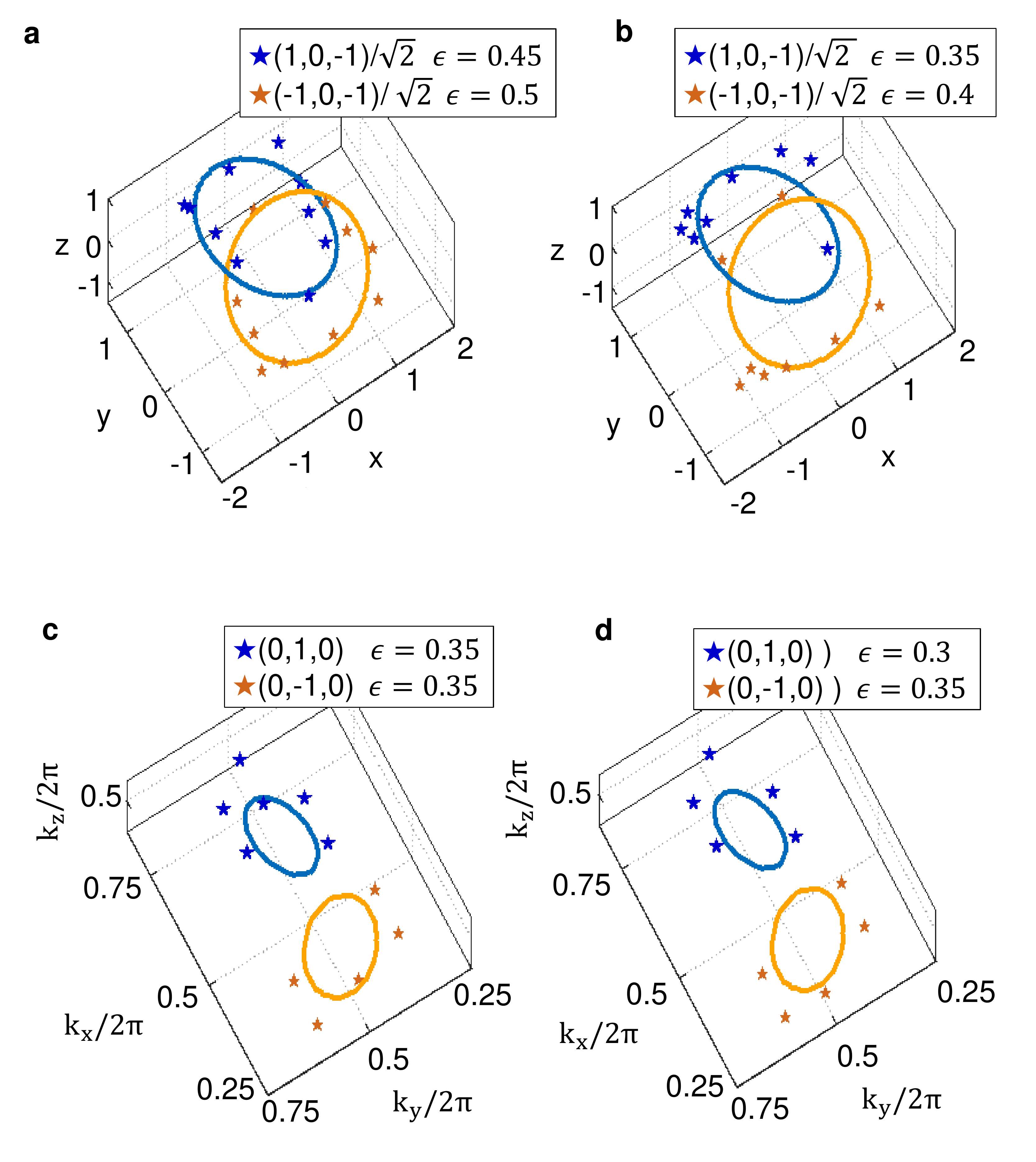}
 \caption{\textbf{The 3D preimage contours that show topological properties of the Hopf insulator.} \textbf{a} Topological link, obtained from two spin states with Bloch sphere representation $\mathbf{S}=(1,0,-1)/\sqrt{2}$ (blue) and $(-1,0,-1)/\sqrt{2}$ (orange), for the legitimate sample with $h=2$ in  the stereographic coordinates. Solid lines are curves from theoretically calculated directions $\mathbf{S_{th}}$ and the stars are experimentally measured spin orientations. The deviation $|\mathbf{S_{exp}}-\mathbf{S_{th}}| \leq 0.45~(0.5)$ for the blue (orange) curve. \textbf{b} Topological link of adversarial examples with $h=2$ with deviation $|\mathbf{S_{exp}}-\mathbf{S_{th}}|\leq 0.35~(0.4)$ for blue (orange) curves. \textbf{c} Unlinked loops of legitimate sample with $h=3.2$.  Two spin orientations are $\mathbf{S}=(0,1,0)$ (blue) and $\mathbf{S}=(0,-1,0)$ (orange). The deviation $|\mathbf{S_{exp}}-\mathbf{S_{th}}| \leq 0.35$ for the blue and orange curve. \textbf{d} Unlinked loops of adversarial example of $h=3.2$. The deviation $|\mathbf{S_{exp}}-\mathbf{S_{th}}| \leq 0.3~(0.35)$ for the blue (orange) curve.  }
 \label{Coutours_in_SM}
\end{figure}

In the main text, we present a one layer's cross section of spin textures. The topological properties of the Hopf insulator are fully captured by the 3D spin texture. The 3D spin textures of the realized legitimate and adversarial examples are shown in Supplementary Figure \ref{allSlice}. This result shows that the spin textures of topological nontrivial phases ($h=0.5$ for $\chi=-2$ and $h=2$ for $\chi=1$) are twisted in a nontrival way, while the topological trivial one ($h=3.2$ for $\chi=0$) is untwisted. The nonzero integer-valued topological invariant guarantees the spin texture cannot be untwisted by continuous deformations. Comparing with corresponding adversarial examples, it is evident that although adversarial perturbations can modify the local details of the spin textures, they do not change its global twisted features and thus the corresponding topological index remains unaltered.

From the Eq. (1) in the main text, we can derive the ground state on the Bloch sphere $\mathbb{S}^2$ with any given momentum point $\mathbf{k}$ in $\mathbb{T}^3$. Consequently, by this equation, one can obtain the preimage contours in $\mathbb{T}^3$ of any spin orientation $\mathbb{S}^2$. To visualize the link and knot for topological nontrivial phases without gluing the boundary, we map the preimage contours in $\mathbb{T}^3$ into $\mathbb{R}^3$. Concretely, the mapping can be decomposited into two parts \cite{yuan2017observation}:
\begin{equation}
\mathbb{T}^3 \stackrel{g}{\longrightarrow} \mathbb{S}^3 \stackrel{s} {\longrightarrow}  \mathbb{R}^3.
\end{equation}
The mapping $g$ first maps the 3D torus $\mathbb{T}^3$ to the stereographic  coordinates $\mathbb{S}^3$ by:
\begin{eqnarray}
\eta_{\uparrow}(\mathbf{k})&=&\sin k_{x}-i\sin k_{y}\nonumber,\\ 
\eta_{\downarrow}(\mathbf{k})&=&\sin k_{z}-i(\cos k_{x}+\cos k_{y}+\cos k_{z}+h).
\end{eqnarray}
where $(\eta_{1},\eta_{2},\eta_{3},\eta_{4})=(\rm{Re}[\eta_{\uparrow}],~\rm{Im}[\eta_{\uparrow}],~\rm{Re}[\eta_{\downarrow}],~\rm{Im}[\eta_{\downarrow}])$ are points on $\mathbb{S}^3$ (up to a trivial normalization). For the purpose of easy visualization, the links are further transformed from the stereographic coordinates of $\mathbb{S}^3$ to $\mathbb{R}^3$ by mapping $s$:
\begin{equation}
(x,y,z)=\frac{1}{1+\eta_{4}}(\eta_{1},\eta_{2},\eta_{3}).
\end{equation}

In Fig.~4 in the main text, we show the 3D preimage contours of topological nontrivial phase $\chi=-2$, which present topological links between two orientations, keep twisted together after adding the perturbations to legitimate samples.
The two orientations of spins in Bloch sphere chosen in the figure are $\mathbf{S_{th}}=(1,0,-1)/\sqrt{2}$ and $(-1,0,-1)/\sqrt{2}$. To obtain the preimage contours on a discrete momentum data grid, we need to select all preimage points with a prescribed tolerance threshold. This can be achieved by defining an $\epsilon$-neighborhood of the desired spin orientation $\mathbf{S}_{\text{th}}$ \cite{Deng2018Probe}:
\begin{equation}
N_\epsilon(\mathbf{S}_{\text{th}})=\{\mathbf{S}(\mathbf{k}):|\mathbf{S}(\mathbf{k})-\mathbf{S}_{\text{th}}|\leq\epsilon\}.
\end{equation}
The choice of $\epsilon$ satisfies condition containing sufficient data points and displaying a clear loop structure simultaneously. We use the same method and obtain preimages in momentum space $\mathbb{T}^3$ for topological trivial phase $\chi=0$ in Supplementary Figure \ref{Coutours_in_SM}\textbf{a} - \textbf{b} and topological nontrivial phase $\chi=1$ in Supplementary Figure \ref{Coutours_in_SM}\textbf{c} - \textbf{d}. From this figure one can clearly figure out that adversarial perturbations do not change the topological links: for
topological nontrivial phases the preimage countours are linked whereas for the trivial ones they are unlinked. 

\end{document}